# Orbital and Physical Characteristics of Meter-scale Impactors from Airburst Observations

P. Brown*[1,2], P. Wiegert[1,2], D. Clark[3] and E. Tagliaferri[4]

[1]Dept. of Physics and Astronomy, University of Western Ontario, London, Ontario, Canada N6A 3K7

[2]Centre for Planetary Science and Exploration, University of Western Ontario, London, Ontario, Canada N6A 5B8

[3]Dept of Earth Sciences, University of Western Ontario, London, Ontario, Canada N6A 5B7

[4]ET Space Systems, 5990 Worth Way, Camarillo, CA, 93012 USA

*Corresponding authors email: pbrown@uwo.ca

**Proposed Running Head** : meter-scale Earth impactors

Editorial Correspondence to :

Dr. Peter Brown
Department of Physics and Astronomy
University of Western Ontario
London, ON
N6A 3K7
CANADA

Phone : 1-519-661-2111 x86458

Fax : 1-519-661-4085

E-mail address : pbrown@uwo.ca

**Abstract**

We have analysed the orbits and ablation characteristics in the atmosphere of 59 earth-impacting fireballs, produced by meteoroids one meter in diameter or larger, described here as meter-scale. Using heights at peak luminosity as a proxy for strength, we determine that there is roughly an order of magnitude spread in strengths of the population of meter-scale impactors at the Earth. We use fireballs producing recovered meteorites and well documented fireballs from ground-based camera networks to calibrate our ablation model interpretation of the observed peak height of luminosity as a function of speed. The orbits and physical strength of these objects are consistent with the majority being asteroidal bodies originating from the inner main asteroid belt. This is in contrast to earlier suggestions by Ceplecha (1994) that the majority of meter-tens of meter sized meteoroids are "…cometary bodies of the weakest known structure". We find a lower limit of ~10-15% of our objects have a possible cometary (Jupiter-Family comet and/or Halley-type comet) origin based on orbital characteristics alone. Only half this number, however, also show evidence for weaker than average structure. Two events, Sumava and USG 20131121, have exceptionally high (relative to the remainder of the population) heights of peak brightness. These are physically most consistent with high microporosity objects, though both were on asteroidal-type orbits. We also find three events, including the Oct 8, 2009 airburst near Sulawesi, Indonesia, which display comparatively low heights of peak brightness, consistent with strong monolithic stones or iron meteoroids. Based on orbital similarity, we find a probable connection among several events in our population with the Taurid meteoroid complex; no other major meteoroid streams show probable linkages to the orbits of our meter-scale population. Our impactors cover almost four orders of magnitude in mass, but no trend in height of peak brightness as a function of mass is evident, suggesting no strong trend in strength with size for meter-scale impactors consistent with the results of Popova et al (2011).

**Key Words:** fireballs, near Earth objects, airbursts.

# 1. Introduction

The nature of the population of meter-scale impacting objects at the Earth is of considerable interest. The flux at meter-scales is responsible for delivery of many of the recovered meteorites at the Earth; indeed almost half of all the known fireball producing meteorites had initial meteoroid diameters in excess of one meter (Borovička et al., 2015). Boslough et al (2015) define an airburst as a bolide with total kinetic energy in excess of 0.1 kT TNT equivalent (where 1 kT TNT =- $4.184 \times 10^{12}$ J). Assuming an ordinary-chondrite-like bulk density (3700 kgm$^{-3}$) and mean impact speed of 20 km/s, this airburst definition corresponds to a size threshold for a spherical rock of one meter diameter. Under these assumptions, this is the size at which meteoroid impacts transition into airbursts as defined by Boslough et al. (2015) and are at the lower end of where ground damage (eg. from an iron meteoroid) might be expected. Such small near-Earth objects (NEOs), which are bodies orbiting the sun having perihelion distances less than 1.3 AU, are also the target population of the proposed Asteroid Redirect Mission (ARM) (Brophy et al., 2012) and therefore of current practical interest.

The origin and composition of NEOs at meter-scales may also provide hard constraints on the delivery and source regions both for meteorites and near-Earth asteroids or NEAs (the sub-population of NEOs which show no extended optical emission beyond a point-source) in general (eg. Bottke et al., 2002); yet this NEA size regime remains poorly studied. There are less than 250 known NEAs having diameters below ~10 m (H > 28) [1] (from a total population estimated to be near $10^8$ (Harris and D.Abramo (2015)) emphasizing the scarcity of data at these small sizes. Physical data on such small NEAs is even less abundant; only ~5% of NEAs in this category have a known rotation period and only one (2008 TC$_3$ – the only asteroid imaged before impact, producing the Almahata Sitta meteorite) has detailed reflectance spectral information (Warner et al., 2009).

---

[1] JPL Horizons – June 10, 2015

Recently, by combining thermophysical and non-gravitational force modeling, and using infrared measurements and high precision astrometry of two small NEAs (2009 BD and 2011 MD), Mommert et al (2014a, 2014b) has been able to place constraining limits on size, bulk density and porosity of small NEAs for the first time. However, such model-based estimates have many free parameters and hence solutions for size/albedo/density are probabilistic in nature for each object. Nevertheless, Mommert (2014b) conclude that the most probable parameters matching the infrared and kinematic behavior of 2011MD suggest it is a rubble-pile assemblage with a most probable bulk density of $600<\rho<1800$ kgm$^{-3}$. This is an important result as it remains unclear what fraction of meter-scale NEAs are rubble-pile assemblages or stronger monoliths (cf. Sanchez and Scheeres, 2014), a question intimately linked to the ultimate origin and evolution of this population.

An alternative method of probing small NEA structure is to observe their interaction with the Earth's atmosphere in the form of bright fireballs. A fireball is any meteor whose apparent peak brightness exceeds Venus (Mv=-4). In this technique, the ablation behavior of the object in the atmosphere provides clues to its internal structure, in particular its crushing strength, as the atmospheric stagnation pressure is assumed to equal the meteoroid crushing strength at points of fragmentation (eg. Baldwin and Shaeffer, 1971). This technique, when applied to the suite of well observed meteorite-producing fireballs (a sample of about two dozen as of 2015) has shown that most such fireballs, while able to produce meteorites with high (tens to hundreds of MPa) compressive strength, are consistent with pre-impact meteoroids having relatively weak (~1 MPa or less) global strength (Popova et al., 2011). This observation has historically been interpreted to suggest most large meteoroids have extensive pre-existing collision-induced cracks (Halliday et al., 1989). None, however, show unambiguous evidence of fragmentation behavior at high altitudes corresponding to the very low-strengths which would be consistent with those expected of a true rubble-pile (where $\sigma < $ ~hPa; Sanchez and Scheeres, 2014). These meteorite producing fireballs, however, are a biased sample of all meter-scale objects colliding with the Earth, as they contained material strong enough to survive as meteorites (eg. Borovička et al., 2015).

Ceplecha (1994) examined all fireball data then available and tabulated 14 fireballs which he estimated to have been produced by meteoroids with pre-atmospheric diameters in excess of 1m. His primary conclusion was that the majority of these meter-scale impactors were weak bodies, likely cometary in nature. This is a surprising result given the lack of observations of a significant population of enduring small (~hundreds of meter) cometary nuclei (Fernandez et al, 2013; Fernandez and Sosa, 2012; Snodgrass et al., 2011) which should be detectable. Cometary fragmentation events do show short-lived small (decameter to hectometer) sized nuclei, but these persist for short periods before disappearing (A'Hearn, 2011). While some authors have claimed telescopic detection of meter to tens of meter-scale objects in meteoroid streams (Smirnov and Barabanov, 1997) these remain unconfirmed (Beech et al., 2004; Micheli and Tholen, 2015), with the lifetime of meter-scale volatile fragments in most major meteoroid streams estimated to be of order only one or a few revolutions (Beech and Nikolova, 2001). Among the NEA population, it is estimated from physical properties and orbital characteristics alone, that 8 ± 5% of the asteroid-like NEO population are cometary in origin (DeMeo and Binzel, 2008) while other studies suggest the fraction may be even lower (eg. Tancredi, 2014). While these values are valid for larger (hundreds of meters to kilometer-scale) NEOs than our meter-scale population, if a true sudden change in population characteristics occurs somewhere in the tens of meters to hundred meter size range, it is potentially very revealing about source populations.

Here we examine the orbits and atmospheric behavior of a suite of 59 fireballs produced by meter-size (diameter > 1m) meteoroids in an effort to constrain their likely physical structure and origin based on ablation behavior and their orbits. As a one meter sized object collides with the Earth roughly once every ~10 days (Brown et al., 2002), large atmospheric area-time products are needed to have a significant prospect of collecting many such events. We make use of three data sources: long running ground-based optical surveys of fireballs, large meteorite producing fireballs and US Government Sensor observations of bolides.

## 2. Datasets and Methods

We isolate meter-scale impactors based on their total preatmospheric kinetic energy. For reference, a spherical one-meter-in-diameter chondritic stone (which we assume has a bulk density near 3700 kg m$^{-3}$ as an upper limit to the observed bulk density of all three ordinary chondrite classes [Britt and Consolmagno (2003)] ) has a mass of ~1900 kg. For objects on typical NEA orbits, Earth impact speeds average ~20 km/s (Morbidelli and Gladman, 1998) which for a one meter chondritic object corresponds to $3.8 \times 10^{11}$ J or equivalently 0.09 kilotons (kT) of TNT equivalent (where 1 kT=4.184 $\times 10^{12}$ J). At the lowest speed for an Earth impactor (11.2 km/s – escape speed from the Earth) the energy is 0.03 kT while at 30 km/s the energy is 0.2 kT. Not knowing the bulk density for most of the events in our dataset (see later) we use this high chondritic density when density is unknown and round our diameter estimate to the nearest first decimal place. This produces a conservative estimate of size, making it unlikely we include smaller objects, but possibly removing a few borderline cases where the bulk density may be lower.

For our data, speeds are usually instrumentally measured with sufficient precision that the speed uncertainty represents a negligible source of error in the estimation of event energy. The determination of the mass is more problematic. In most cases, the radiation emitted by the bolide is summed over time and some integral luminous efficiency, $\tau$, must be used to estimate true mass (and hence total energy) (see Borovička et al., 2015 for a detailed physical description). As summarized in Nemtchinov et al (1999), earlier estimates of $\tau$ (eg. Ceplecha and McCrosky, 1976) were based on experimental data from small (gram-sized) artificial meteoroids, resulting in $\tau < 1\%$ at speeds below 30 km/s. For meter-scale objects, the radiation emission becomes particularly complicated – theoretical estimates of $\tau$ for large, deeply penetrating bolides (in differing passbands) have been estimated by Nemtchinov et al (1997), Golub (1996) and from experimental fits to fireball data by ReVelle and Ceplecha (2001). All of these studies suggest values of $\tau$ are several times larger than earlier estimates. By using both larger values for $\tau$ and explicitly incorporating the role of fragmentation, Ceplecha and ReVelle (2005) were able to reconcile the long standing problem of large (one to two order of magnitude) differences between fireball masses computed from radiation (photometric masses) as compared to masses computed from deceleration (dynamic mass).

For meter-scale bodies, independent confirmation of the theoretical estimates for τ produced by Nemtchinov et al (1997) were made by Brown et al (2002) and later Ens et al (2012) who compared empirical energy estimates for meter-scale impactors derived from infrasonic periods with optical energies of the same events measured by US Government sensors (Tagliaferri et al., 1994). The resulting integral τ values in the passband of the optical detection systems range within a factor of two approximately from 5%-10% for speeds of 12 – 30 km/s for meter-scale impactors. On theoretical grounds tau is expected to increase with speed and mass (Golub et al., 1996).

*2.1 Ground-based fireball networks*

Several camera networks have been established and operated for decades or more with the express purpose of photographing meteorite producing fireballs (Oberst et al., 1998). The three principal networks include:

1. US Prairie Network (PN) which operated between 1963-1975 in the midwestern states of the USA (McCrosky et al., 1976; Ceplecha, 1986)

2. the European Fireball Network (EN or EFN) which began regular monitoring of the sky using a network of cameras in 1963 and continues in operation to the present day (Oberst et al., 1998; Spurný et al., 2007)

3. The Meteorite Observation and Recovery Project (MORP) which operated from 1971-1985 in the Canadian prairies (Halliday et al., 1978; Halliday et al., 1996).

To estimate the number of meter-scale objects expected to be detected we need the effective atmospheric collecting area time product of these networks. As a baseline, using the average global energy – flux values for meter-sized impactors (assuming 0.09 kT corresponds to a meter-sized body) given in Brown et al (2002) we find the mean time-area product for such an impact to be ~$10^{11}$ km$^2$h. For the MORP clear-sky survey, the area-time product was $1.5 \times 10^{10}$ km$^2$h (Halliday et al, 1996) and of order ~$10^{11}$ km$^2$h for the entire operating time of the PN based on the yearly average given in McCrosky and Ceplecha (1968) for the brightest events (roughly corresponding to meter-scale impactors). Large fireballs captured by these networks typically have high precision astrometric and photometric records permitting in-depth modeling of the ablation and fragmentation of the meteoroid facilitating inferences about its physical structure (see Borovička and Spurný (1996) for an example of such a study).

There is no equivalent time-area published for the entire operating time of the EN, which has changed in area, scope and instrumentation several times over the last 50 years. Oberst et al (1998) estimated the early temporal and spatial coverage of the EN to be comparable to MORP/PN (which averaged near $3 \times 10^9$ km$^2$h per year). Using this average yearly time-area product and accounting for changes in the network in the late 1990s and early 2000s (recognizing this is a rough approximation as there is considerable variation in the last few decades as the network has changed area and equipment) and the fact that very bright fireballs are often recognized in poor sky conditions at long ranges in the network, we crudely estimate an integrated time-area product for multi-station coverage of clear skies to be of the order of ~$3 \times 10^{11}$ km$^2$h for the EN for meter-scale events to the end of 2014. On the basis of these network collecting-area time products, we predict that the MORP clear sky survey would have only a 15% probability of detecting a meter-sized or larger impactor, so zero detections as observed is expected.. Based on their time-collecting area products we expect the PN would have had a 65% chance of recording one or more meter-sized or larger impactor and the EN a 95% probability of detecting one or more such impactors .This will be revisited in more detail in the discussion section.

Using the more recent values for $\tau$ to estimate photometric masses, we find no meter-class impactors detected by MORP, one by the PN and 4 by the EN, one of which (Benesov) produced recovered meteorites (and is shown only in Table 2). Note that for MORP, Halliday et al (1996) have consistently used a fixed $\tau$ of 4% and we have used their photometric masses as reported to infer size. For the EN and PN data, masses were derived from recent re-analyses of these events using larger $\tau$ as given in Ceplecha and ReVelle (2005) and Nemtchinov (1996). Table 1 summarizes the orbital and physical properties of the three meter-scale camera network events which did not produce recovered meteorites. We note that some of the earlier (pre-1970s) EN events have not been fully re-analysed with more recent values of $\tau$, but we may estimate the "modern" equivalent photometric masses approximately knowing that the original masses were based on the Ceplecha and McCrosky (1976) luminous efficiency scale. From this comparison we have concluded that none exceeded meter-sizes. This approach excludes 2-3 events which might potentially be border-line cases, a small number relative to our total sample size.

## 2.2 Meteorite-producing fireballs

As of early 2015, some 23 meteorite falls have been instrumentally recorded with sufficient accuracy to allow pre-atmospheric orbits to be computed (Borovička et al., 2015; Trigo-Rodriguez et al, 2015). Among these, ten had pre-atmospheric masses placing them at or above our one meter diameter threshold. Table 2 presents data for these events as determined by the associated references. One of these (the Benesov meteorite fall – EN 070591) was also recorded by multiple stations in a regular fireball survey (EN) and meteorites subsequently recovered (Spurný et al., 2015). For clarity we put this event only in Table 2. Here we assume the measured bulk density (where available) of the recovered meteorites is the same as the original meteoroid bulk density together with the nominal initial mass estimates from each study to estimate pre-atmospheric size. This moves the Maribo fall to just above the one meter limit while Moravka falls below our size limit. Taking into account the actual mass uncertainty range, both of these falls might be just under or over our nominal one meter limit. Similarly, the recent revision of the Pribram initial mass downward (Borovička and Kalenda, 2003) using modern values of $\tau$ and fragmentation nominally removes it from our list. The other meteorite producing fireball events given in Table 2 are confidently meter-class or larger.

While these events have been observed by varying instruments, potentially producing scatter due to systematic errors, the ground-truth provided by recovered meteorites and the (often) multiple independent techniques used to estimate or constrain initial mass make these a unique calibration dataset (see Popova et al (2011) and Borovicka et al (2015) for a detailed synopsis of some of these events and their associated mass measurements).

## 2.3 US Government sensor detections

The majority of our dataset (47/59) consist of fireball detections reported from US Government (USG) sensors. These data appear online at the NASA JPL fireball and bolide reporting website2 and we use the terminology given on that website for the data (calling these US Government sensor data throughout). As described on the website (and references therein), the equivalent radiated energy is computed assuming 6000K blackbody emission (see Tagliaferri et al., 1994; Brown et al., 1995; Nemtchinov et al, 1997 and Brown et al 2002 for more details). This radiated energy is then converted to a total initial kinetic energy assuming an integral τ as empirically determined in Brown et al (2002).

As of the end of April, 2015 a total of 76 events are reported on the site. For our study we selected the 50 events which have a radiant and speed provided, a height of maximum luminosity and a total estimated energy. Finally, we required the initial impactor diameter to be larger than 1m using a nominal bulk density of 3700 kgm$^{-3}$ together with the known mass (from speed and energy); this further removed 3 events, leaving us with 47 meter-scale impactors. Note that one of these events, Chelyabinsk, is already shown in Table 2. The remaining 46 events and their associated orbits are given in Table 3.

Orbital elements and their uncertainty were computed assuming the last significant figure given in the online JPL table for the velocity component and geographic location represents the precision of the measurement. The corresponding uncertainty in the velocity dominates the orbital error and is assumed to be of order 0.1-0.2 km/s. Orbits were independently computed following the monte carlo numerical technique of Clark and Wiegert (2011) as well as the analytical approach described in Ceplecha et al (1987) as a cross-check. The resulting orbital uncertainties were found by generating 1000 clones from the initial state vector within its uncertainty and following these orbits numerically backward for 60 days to estimate the associated orbit error at infinity. Median uncertainties are 2% in semi-major axis, inclination and eccentricity.

---

2 http://neo.jpl.nasa.gov/fireballs/

The uncertainty in measured peak height is more difficult to estimate from the available data. However, seven fireball events with heights of peak luminosity reported by USG sensors also have independent estimates from ground-based data. These events are summarized in Table 4. In general, most height estimates appear to agree to within a few kilometers, the average difference being 3 km, which corresponds to approximately a factor of 1.5 in equivalent dynamic pressure. The largest difference occurs for the Chelyabinsk fireball, possibly connected with the complex fragmentation and debris cloud formed during this exceptional event (eg. Popova et al (2013)).

The accuracy of the radiant and speed for USG events can be gauged by comparing trajectory data computed for the four events in Table 2 for which independent ground-based estimates and USG data are simultaneously available. These four meteorite-producing fireball events (Almahata Sitta, Buzzard Coulee, Chelyabinsk and Kosice) have USG trajectories and equivalent orbits shown in bold in Table 2. Three of the four have speeds within a few tenths of a km/s, while two (Kosice and Chelyabinsk) also have radiants within ~2 degrees of the independent ground-estimates.

Buzzard Coulee (BC) shows a large difference between the ground-based determined speed (18 km/s) and USG speed of 12.9 km/s. Similarly, the radiant measured from USG data and from ground-based observations differ by more than 45 degrees. BC is also the least precise of these four ground-based events (Borovicka et al (2015)), having a trajectory reconstructed from indirect shadow measurements (Milley, 2010), so the significance of these differences is unclear.

In contrast, the almost 30 degree difference in radiant position between the USG measurement and the known radiant for Almahata Sitta (2008 TC3) [AS] is much more significant. AS is the most precisely determined orbit and radiant of any meteorite-producing fireball, having been detected in space almost one full day prior to impact (Jenniskens et al., 2009). The speed difference is only 0.9 km/s and as a result the actual errors in the orbital elements are modest, but this case emphasizes that USG radiant determination may be problematic in some instances.

The detection biases for USG data are not available. However, it is notable that of the events with peak brightness altitude listed, the fraction with velocities also provided is 75% and constant with altitude for heights below 40 km. In contrast, only half of the events with altitudes over 40 km have velocities reported. This suggests the USG data may have a selection bias against higher (>40 km) altitudes and hence higher speeds on average. This in turn may reflect a bias against faster (cometary-type) objects. We suggest that the population fractions of cometary objects described later may be best interpreted as lower limits.

## 3. Results

### *3.1 Orbit associations*

As discussed in Gounelle et al (2008) and refined in Tancredi (2014), the main parameter in the modern orbital-dynamics-based classification system for comets and asteroids, is the Tisserand parameter with respect to Jupiter ($T_j$). In this system, objects having orbits with $T_j > 3$ are decoupled from Jupiter, while those with $2 < T_j < 3$ cross Jupiter's orbit and have their subsequent dynamical evolution driven by encounters with the planet. The former are generally main-belt asteroids and the latter are typically Jupiter Family comets (JFC), though in the interval $3 > T_j > 3.05$ Jupiter may still strongly affect an orbit, so JFCs arguably extend into this range of $T_j$ (see Tancredi (2014)). When $T_j < 2$ and a < 40 AU objects are defined as Halley-type comets (Levison, 1996). Note also that there is mixing of main-belt asteroids below $T_j < 3$ as discussed in Bottke et al (2002), so this criterion alone is not sufficient to uniquely identify objects as definitively JFC in origin.

The orbits given in Tables 1-3 were compared with all known NEA and comet orbits from the JPL Horizons web-site[3] and meteor shower orbits from the IAU Meteor Data Centre[4]. The algorithms used were the standard orbit dissimilarity criteria proposed by Drummond (1981). Additionally, the false-positive probability between our test orbits and the NEA population was computed following the technique outlined in Wiegert and Brown (2005) and updated in Borovička et al (2013) to include parent-body size weightings.

Streams of meter-scale and larger NEOs might be created via several mechanisms such as YORP spin-up disruption, fragmentation from impact, tidal disruption during previous close Earth passages or cometary disintegration. We note that the decoherence timescale for meteoroids of meter-scale is of order $10^5$-$10^6$ years (Pauls and Gladman, 2005).

We find no significant linkages with any comets, an unsurprising finding as only ~10 of our orbits are on JFC/HTC-type orbits having $T_j<3$.

---

[3] Accessed on April 20, 2015
[4] Accessed on April 20, 2015

To test shower linkages, we pick initial loose values of the D' (Drummond D) cutoff of 0.05 (see Jopek and Froeschle, 1997 for a discussion of thresholds given dataset size). Our reasoning is that if no meteoroid stream orbit is a match with this initial permissive value, it is unlikely to be a real association. Note that we are comparing 59 fireball orbits with ~500 meteoroid stream orbits.

We find only three fireballs among all our dataset which potentially match any shower following this definition. The first, the Sumava fireball, (EN 041274), is linked to the N. Chi Orionid stream (ORN) (IAU stream number 256) with D'=0.04. This shower linkage with Sumava has been recognized earlier (Ceplecha, 1976; Borovička and Spurný, 1996). The N. Chi Orionids are commonly interpreted as linked to the broader Taurid meteoroid complex (Porubcan and Stohl, 1987), which has previously been identified as unique among meteor showers in producing large fireballs (eg. Wetherill, 1974). We note that the shower radiant is located in the densely populated anti-helion sporadic meteor source and hence non-Taurid contamination is possible (Brown et al., 2013a). Finally, the mean orbital elements for the diffuse stream differ among literature sources (see Jenniskens, 2006 for a summary), making a definitive association questionable.

The second linkage is between the Maribo meteorite fall and IAU shower #96 (NCC), the Northern Delta Cancrids also with D'=0.04. As with the ORN, the NCC shower radiant lies in the anti-helion sporadic source and has also been linked to the Taurid complex of streams (Jenniskens, 2006). Various surveys list differing orbital elements for the shower (Kronk, 2014) so this association also remains tentative.

The final possible match is PN 39470 with the Southern Chi Orionids (IAU shower #257) with D'=0.05. This shower is also a potential sub-stream of the Taurid complex (Jenniskens, 2006).

It is notable that of the foregoing showers only the S. Chi Orionids is listed as an established meteor shower by the IAU; the other two are among the working list (showers which are yet to be confirmed) of ~500 streams. We find no compelling evidence of any established, major meteor shower signature among our dataset of meter-scale objects, a conclusion consistent with Schunova et al (2012) who found no clear clustering among orbits for NEAs. All three of our potential shower linkages are with possible sub-streams of the Taurid shower, the only shower complex where large meteoroids have been definitively observed (Jenniskens, 2006; Brown et al 2013a).

To compare our impactor orbits with NEA orbits we use the false-positive association criteria described in Wiegert and Brown (2005). Synthetic NEAs are drawn from the Bottke et al (2002) debiased NEA distribution until a better match (lower D') is found. The number of draws n required provides an estimate of the probability that any asteroid drawn randomly from the NEA population would provide as good a match or better, that probability being $1/n$. If the number of asteroids which are at least as large as the hypothesized parent is N (calculated from the NEOWISE estimate of the NEA population, Mainzer et al 2011), then the ratio $N/n$ provides a measure of the significance of the association: $N/n \ll 1$ implies coincidental similarity is unlikely. High statistical significance does not of course prove a physical/genetic relationship, but this procedure does provide measure of significance against the local background distribution. In regions where NEAs are abundant ie. at low inclinations, a better match between hypothesized parent and bolide is needed to achieve the same significance than it is where NEAs are less numerous ie. at high inclinations.

Note also that this is an explicitly size-dependent criterion, as N is smaller for larger hypothesized parents. This is justified here on the basis that 1) associations of bolides with larger parent asteroids are more interesting (as more mass is available from the parent so any breakup/decay mechanism produces more fragments of a given mass) and 2) we see no other clear-cut way to define the sample size, which is open-ended and progressively more observationally biased towards smaller sizes.

We also compare the cumulative D' distributions for NEAs near the proposed parent to scaling laws which are expected to follow $D^{4-5}$. Since orbital similarity criteria such as D compare 5 of the 6 orbital parameters, a uniform distribution around them should grow like $D^5$. NEA distributions typically deviate from this due to observational biases, which remove one degree of freedom and drive the distribution towards $D^4$. Studies of the NEA distribution by Reddy et al (2015) have shown that the exponent is approximately 4.4. If the closeness of the association between the hypothesized parent and the bolide is beyond that which can be explained by random association with the asteroidal background, the cumulative D plot will lie to the right of the $D^{4.4}$ line; this is seen in the cases discussed below.

Table 5 summarizes our most significant NEA-fireball associations, where the orbit dissimilarity Monte Carlo and size-weighting procedure suggests an association at the <1% level. Ten possible associations are produced in this way, but we note that almost half are just barely beyond the 1% (arbitrary) threshold and many of these are with smaller NEAs where an association is likely to be less physically plausible. We discuss only a subset of the more interesting possible associations.

By far our most significant NEA linkage is between Chelyabinsk and 86039, with a <0.01% chance of the D' of 0.018 being purely chance for such a large object. That 86039 has H=16 makes this a much more interesting possible connection. This relationship has been suggested previously by Borovička et al (2013) based on the orbital similarity. More recently, Reddy et al (2015) have questioned this link based on the dynamical difficulty of such a relationship (ie. it would have to be very young, which is improbable and the mechanism for formation is nebulous) and the dissimilarity in the reflectance spectra between asteroid 86039 and recovered Chelyabinsk meteorites.

A less significant match is found between the 1.1 km diameter 85182 (1991 AQ) and the Maribo meteorite fall. At D'<0.044 this also appears to be significant relative to the local density of known NEAs with H<20 as shown in fig 1. However, the spectral class of 1991 AQ (Sq) [Somers et al (2010)] is not consistent with a CM2 chondrite, so this may be a chance association. The difference between the surface and bulk properties of the asteroid or internal heterogeneity in the asteroid may also play a role.

The NEA 2001 UA5 (163014) is a roughly 1-2 km body which has a D'=0.033 with USG 20121120 and a resulting false positive level of slightly less than 1%. Binzel et al (2004) found 163014 to be spectral type Sq. The significance of this connection is less clear as no meteorite was recovered from this fireball; its peak brightness and speed are consistent with a type II fireball similar to Maribo (see next section).

The final NEA-fireball orbit linkage of note is between PN 39470 and 2201 Oljato. This linkage is interesting both because the D' of 0.043 is low, the probability of random association with an NEA of this size at this level is 1 in 250 and 2201 has a diameter of ~4 km. Moreover, 2201 Oljato has been long proposed as a parent body for both dust trails (Lai et al 2014), meteoroid streams (Drummond, 1981) and a probable extinct cometary candidate based on excess UV emission (McFadden et al, 1993). The cumulative D-plot (Figure 1) shows that there are relatively more objects in similar orbits compared to the earlier three candidates, reflecting the dense population of NEAs in the general region of the Taurid stream, though all are much smaller than 2201.

*3.2 Orbit distribution*

Figure 2 shows the aei orbital distribution of our observed meter-scale impactors. There are no high inclination orbits in our data; indeed only one object (USG 20091121) has an inclination above 35°, an indication of the lack of long-period comet signature in the meter-scale impactor dataset as long-period comets have an isotropic inclination distribution, while NEAs have inclinations which peak near the ecliptic. .

The overall shape of the aei data resembles that predicted by Veres et al (2009) based on the Bottke et al (2002) NEO model, weighted by collision probability (see their figure 1). The mode of their distribution is near a~1 and e~0.5, matching our observed distribution. Our inclination distribution is slightly broader, but given our small number statistics the difference is unlikely to be significant. Using the orbital elements for all 59 events we may also apply the Bottke et al (2002) NEO model to estimate their original source regions. The Bottke model assumes the orbital elements for the present NEO population is size independent; this is likely a poor assumption at our sizes where the dynamics are likely influences by the Yarkovsky effect, (eg. O'Brien and Greenberg, 2005) and even meteoroid impacts (Wiegert, 2015). With this caveat, in the absence of a more comprehensive delivery model for such small objects, we apply results from the Bottke et al (2002) NEO model to our orbital dataset.

Figure 3 shows the cumulative normalized probability distribution in terms of the original Bottke five source model (see the Figure 3 caption for a summary of these sources) applied to our impactor data. In this figure the probabilities for 55 events have been summed for each source and then normalized. Note that as the Bottke model does not include an HTC comet source, our four events with $T_j<2$ are not included in this figure. We find that the $\nu_6$ resonance source dominates the delivery for our meter-scale impactor population with ~50% of the probability originating near the inner main asteroid belt, by far the highest of all five source regions. This is similar to the 46% found by Binzel et al (2004) for the source of ~400 NEOs with known spectral taxonomies, but higher than the 37% prediction of the proportion of NEAs from the $\nu_6$ in the steady-state population by Bottke et al (2002) or the 41% predicted by Greenstreet et al (2012) for the Earth-impacting population.

As our number statistics are small (and the applicability of the Bottke model to our small sizes questionable) we nevertheless consider the agreement suggestive of a common source population, particularly if one considers that the overall source region percentage for delivery from the $\nu_6$ across all 59 meter-scale impactors is actually 46% (7% being from HTCs). Table 6 compares our source probabilities with those found by Binzel et al (2004) for NEAs having taxonomic classification, with the Bottke et al (2002) steady-state values per source and the Greenstreet et al (2012) source probabilities for NEOs impacting Earth. Our largest deviation from these sources is with the 3:1 and $\nu_6$ but given our statistics (and the lack of Yarkovsky effects in the models) we consider the global agreement is quite good.

Our lower limit for the total "orbital" cometary fraction (that is impactors which have either HTC or JFC-type orbits) is 15%. This suggests that 85% - 90% of all meter-scale impactors at Earth are derived from the main-asteroid belt, with a ~5%-10% JFC fraction among NEA-type orbits, based on orbital classification alone. Several independent estimates place the fraction of NEOs with $2<T_j<3$ also having low-albedos consistent with cometary nuclei as high as ~50% (Fernandez et al., 2005; DeMeo and Binzel, 2008; Mainzer et al, 2012), which translates into the fraction of all NEAs which may be extinct JFCs as ~10%, in good agreement with our values. Note, however, that these studies are applicable to ~km-sized NEAs while the albedo distribution of cometary fragments at meter-scales is unmeasured.

Interestingly, our impact-derived HTC fraction of 7% is comparable to our JFC fraction. The lack of HTC telescopic detections with q < 1.3 AU (only 1 Damocloid with q<1.3 AU and a well determined orbit was known prior to 2007) has led to this source being excluded from NEO models for lack of statistics (see the discussion in Greenstreet et al (2012)). Our work suggests that at small sizes, the HTC impactor component may not be negligible. Moreover, given the potential biases discussed in section 2.3 for USG sensor detections at high altitudes/large speeds, this is a lower limit.

Figure 4 shows the distribution of impact speeds for our dataset. Our average impact speed and standard error is 18.3 ± 0.7 km/s while our median speed is 17.6 km/s, roughly intermediate between the $V_{rms}$ impact speed of 17 km/s suggested by Chyba et al (1994) and the 20.6 km/s predicted by the Greenstreet et al (2012) NEA model. We note a lack of any significant high speed (>40 km/s) large (meter-class) fireballs, despite the fact that such meter-scale impactors would produce very energetic airbursts per unit mass. This either represents a large-scale bias in the USG sensor data or further strengthens the view that meter-scale objects are rare among the high speed HTC/NIC population where very weak cometary material might be expected.

We also find that the mean impact angle for our dataset is $46° \pm 3°$, consistent with the often quoted expected average impact angle of 45° (Melosh, 1987).

### *3.3 Physical structure*

The ablation behavior of meteoroids (deceleration, light curve, end height, flares etc.) provides clues to their physical structure (see Borovička (2006) for a recent review).

Previous analyses of fireball data have strongly suggested that meteorites are produced from meteoroids which are substantially weaker than their recovered fragments (Halliday et al., 1989; Popova et al 2011; Borovička et al 2015). Popova et al (2011) analysed all fireball producing meteorites then available and concluded that the parent meteoroids were typically weak with global strengths of order 0.1 – 1 MPa. Peak strengths at the later (deeper) major breakup were found to be an order of magnitude higher. This leads to a modern picture (eg. Borovička et al 2015) wherein most meter-scale objects are permeated by large scale fractures (in the case of ordinary chondrite-like material) or possibly have high micro-porosity (ie. carbonaceous chondrites) resulting in structurally weak bodies. Exactly where in the structural spectrum meter-class meteoroids reside, whether as coherent rubble-piles or aggregates as defined for asteroids (Richardson et al 2002) is more difficult to determine as we have only fragmentation behavior and a light curve to forensically reassemble the original object. The resulting inversion is not unique.

The modern picture of the structure of meteoroids producing fireballs is built on the work of Ceplecha and McCrosky (1976), (hereafter CM76) who examined some 200 Prairie Network (PN) fireballs (ranging in size from ~cm to sub-meter-scale initial diameters) to identify characteristics of each event which might be used as a diagnostic for material strength and structure. They noted that meteoroids having similar masses, velocities and entry angles often reached end heights (the final altitude at which the fireball was luminous) which were very different. These end height differences could be directly interpreted as strength differences, presuming that the aerodynamic ram pressure ($\rho v^2$) controlled fragmentation, a dominant process in meteoroid ablation (Ceplecha et al., 1998). As a result of their analysis, they proposed a one dimensional criterion (still widely used today) termed PE which can be used to gauge the relative strength of meteoroids producing fireballs based on a statistical fit of PN data which included the initial mass (based on photometric masses), entry velocity and entry angle.

More specifically, as originally given in CM76, the PE value is given by:

$$PE = \log(\rho_E) - 0.42\log(m_\infty) + 1.49\log(V_\infty) - 1.29\log(\cos Z_R) \qquad [1]$$

where $\rho_E$ is the atmospheric mass density (in units of g cm$^{-3}$) at the height of the fireball end point (and also is the origin for the acronym PE ($\rho_E$) emphasizing the physical link between body strength and atmospheric mass density at the end height), $m_\infty$ is the original meteoroid mass in grams, computed from the total light production, $V_\infty$ is the entry speed in km/s and $Z_R$ is the local entry angle from the zenith. Larger PE values (less negative) correspond to stronger material displaying less ablation. CM76 further proposed that specific "types" of fireballs might be distinguished as four distinct taxonomic classes via the PE criteria. These strength groups and their probable material association are given as (Ceplecha et al. 1998):

type I :   PE > -4.60                Ordinary Chondrite-like

type II:   -5.25 < PE ≤ -4.6         Carbonaceous Chondrite (CI/CM)

type IIIa: -5.7 < PE ≤ -5.25         Short period cometary

type IIIb: PE ≤ -5.7                 Weak cometary material

ReVelle and Ceplecha (1994) also proposed a separate "type 0" for iron meteoroids, though the evidence for this population at fireball sizes is limited. Note particularly that the association of type I fireballs with ordinary chondrites (or comparable-strength stony meteorite material) has been largely validated by recorded meteorite falls [see Table 2 and smaller events discussed in Popova et al. (2011)], but that mixing of comparable material between adjacent types certainly occurs. The PE limits should not be strictly interpreted to correspond rigorously with the physical character of meteoroids as emphasized in CM76. Furthermore, note that the mass scale used for [1] is appropriate to the original scale given in CM76, which used (by modern standards) low values for luminous efficiency (and hence the photometric masses were typically too high). Nevertheless, the general (relative) trend is expected to remain valid; for a uniform mass scale, smaller PE values represent weaker material.

Examination of Table 2 shows that for the ten meteorite falls produced from meter-scale progenitor meteoroids with known end-height, an estimate for the PE value (and associated fireball-type) can be made – this is shown in the third to last column of Table 2.

The fireball types are derived from the original references or the stated end heights and masses; we caution again that these masses (which are more likely "true" masses as they are often found from several independent means) are likely smaller than expected from the CM76 luminous efficiency values so there may be a systematic offset toward larger values of PE. Nevertheless, it is reassuring that for these cases the majority of the ordinary chondrite events fall in the fireball type I category as CM76 predicted. Kosice and Almahata Sitta (2008 TC$_3$) are exceptions - in both cases detailed analyses have shown these to be atypically weak objects [Borovička et al. (2015)], while the three carbonaceous chondrite falls are associated with fireballs of type II or IIIa categories. Chelyabinsk nominally falls in the type II category, but the energy and size of this object is so much larger than the data comprising the original CM76 dataset it is unsurprising that in such a huge extrapolation the CM76 system breaks down. Thus to the extent that the events in Table 2 represent calibrated (or ground-truth) comparable in size to the original dataset of CM76 fireballs, we have some confidence that in broad form the interpretation of PE is valid.

One major limitation in our data, however, is that all US Government sensor reported bolides do not have reported end heights, but rather only heights of maximum brightness. The height of maximum brightness is determined by a complex interplay between fragmentation heights, material properties and speed (see Ceplecha and ReVelle, 2005 for examples of different possible light curve shapes depending on fragmentation). Nevertheless, we expect stronger material, more resistant to fragmentation, to generally have peak brightness lower in the atmosphere. An exception to this picture is expected for very large airbursts where forward momentum may carry material downward lowering the effective burst height (Boslough and Crawford, 2007).

In Figure 5 we plot the height of maximum brightness for all our fireballs produced from meter-scale objects as a function of speed. Also shown are the heights of peak brightness for nine of the meteorite producing fireballs (Table 2) and the three fireball network events (Table 1) which have well determined PE values and therefore can be assigned fireball type on the basis of PE. The actual luminous end heights are all below the heights shown in Figure 5, though for weaker objects the height of peak brightness and end height are very similar (Sekanina, 1983). Also shown are contours of equivalent aerodynamic load as a function of speed and height. In most cases, we expect the actual strengths of the objects to be lower than the pressure at their peak luminosity heights – ie they have undergone fragmentation higher in the atmosphere than the height of peak brightness. In this sense, the pressure contours may be thought of as extreme upper limits to the global strength of the original meteoroid.

As discussed in Popova et al (2011) and Borovička et al (2015), for most well-observed fireball-producing meteorites, the initial fragmentation occurs under pressure loads 1-2 orders of magnitude lower than the compressive strength of the associated meteorites. Typical values are 0.1 – 1 MPa at initial fragmentation, with some fragments surviving pressures of a few to 10 MPa before fragmenting further. Typical ordinary chondrite meteorite compressive strength values measured in the lab are tens to a few hundred MPa as summarized in Popova et al (2011).

We expect stronger objects to reach peak brightness lower in the atmosphere, though variations in strength and fragmentation behavior restrict this to a statistical expectation.

There is also a weak trend of higher peak heights as entry angles become more shallow and hence peak brightness height as a criterion for global strength is likely a poor measure for objects having nearly horizontal entry angles. We are reluctant to try and associate specific quantitative measures along the lines of the PE criterion to our small dataset simply because of this expected variability. However, one advantage of using the height of peak luminosity as a structural diagnostic is that it is more representative of the nature of the entire body, being determined by the overall fragmentation behavior. In contrast, the end height is often representative of the strongest material in the body as a whole. The absolute end height is also a measurement more subject to biases created by instrument sensitivity, affected by range and local cloud conditions. In contrast, the height of peak brightness is usually well defined for both bright and fainter events.

To help guide interpretation of the observations in Figure 5, we also have plotted the modeled height of peak brightness for the CM76 fireball classes using the Triggered Progressive Fragmentation Model (TPFM) described in ReVelle (2005). This model has been previously applied successfully to a number of meteorite producing fireballs in our dataset including Benesov, Tagish Lake, Park Forest (ReVelle, 2007) and Chelyabinsk (Brown et al., 2013b). As our data are an energy limited sample, for the model we use the median energy of our population (0.4 kT), together with the fireball type properties (ablation parameter, tensile strength, assumed bulk density), self-similar shape-change parameter (Mu) of 0.667 as described in ReVelle (2005) with modifications to the fragment cascade strength in the model as proposed in Brown et al (2013b), whereby each subsequent fragmentation occurs at 1.3 times the value of the previous episode. Details are given in the caption to Figure 5. We note that it is the strength and assumed fragmentation behavior which control the vertical positioning of the blue (model) lines.

We find that the four photographic network fireballs (triangles in Figure 5 plus the Benesov meteorite fall) in the meter-sizes class with known PE fall are consistent with the TPFM modelled fireball type class using their peak brightness/speed alone. It is further reassuring that the meteorite-producing events, where we have both knowledge of the nature of the meteoroid, are consistent with the model trends from TPFM shown in the figure. It is also clear from Figure 5 that the calibrated meteorite-producing fireball events fall in the middle of the field of all other data; this suggests that the majority of the meter-scale meteoroids ablate similarly to the well documented fireballs producing recovered meteorites in terms of strength/fragmentation characteristics.

About ten events are near the border between type II/IIIa – this is also where the meteorite-producing fireballs Almahata Sitta (AS) [2008 TC$_3$] and Kosice (K) reside. Both were found to be unusually weak bodies (see the discussion in Borovička et al (2015)).

The case of Almahata Sitta (2008 TC$_3$) is particularly unusual – the object was clearly globally quite weak/fragile. Some studies suggesting it had an initial porosity as high as 50% (eg. Kohout et al., 2011), though whether it was a true rubble-pile or simply a weak body (Borovička et al 2015) remains unclear. Borovička and Charvat (2009) noted that several early flares before the main flare at 37 km altitude for AS confirms its weak nature. The same authors in trying to explain the high altitude fragmentation of AS applied a fragmentation destruction model, which had previously been successful in explaining the extremely weak Sumava fireball (which appears in Fig. 5 in the extreme upper right hand side of the diagram at a velocity of 25 km/s and H$_{peak}$ of 67 km) and predicted that had AS had similar physical properties to Sumava it would have produced a terminal flare around 55 km altitude at the AS speed of 12 km/s. This result is interesting as at a similar (low) speed to AS the USG event of Nov 21, 2013 (USG 20131121) showed a peak brightness at 60 km altitude, strongly suggestive of a very fragile body fragmenting into many small pieces. This event is isolated from the other events in Figure 5 and the only one of the USG dataset which falls in the TPFM category of very weak (type IIIIb) cometary material. It is most similar in behavior to the Sumava fireball (Borovička and Spurný, 1996) which is the only well documented meter-scale fireball showing unquestionably extremely weak (cometary) fragmentation behavior. The pre-impact orbit for the 2m diameter USG 20131121 is Aten-class, a highly evolved orbit perhaps an indication of the long collisional timescale for such a large object.

Most other events at low speeds comparable to AS show much lower heights of peak brightness suggesting the majority are stronger objects (type II or I) as compared to AS and/or Kosice. Of the four other USG events in TPFM category IIIa (or borderline II/IIIa), two show typical Apollo-type orbits (USG 20120826 and 20120421) while another (USG 20140516) is an Aten. The fourth fireball (USG 20150102 – located at 18,38) is one of only four events having orbits with $T_j < 2.5$. Together with USG 20150107 (located at 35, 46) these are the best candidates among the USG events for meter-scale cometary fragments, having both a cometary-type orbit and atmospheric behavior consistent with an object on the weak end of the strength spectrum relative to our other events. The strength, however, may be most similar to carbonaceous chondrites, as both fall in the TPFM type II field and are not very different from SM, a known CM2 fall.

In figure 5 the associated orbits are color-coded by Tisserand parameter following the classification of Levison (1996). The majority of orbits are asteroidal ($T_j>3$) (50/59); of the remaining 9 orbits, four are nominally in HTC-type orbits (none showing high inclinations) while five show JFC-type orbits. Note that two of the latter are Maribo and Sutter's Mill which both produced CM2 carbonaceous chondrites and are among our calibrated (meteorite-producing) fireball dataset. Both also have $T_j \sim 3$. They are most likely derived from the main-belt (see discussion in Jenniskens et al (2012), though an association with JFCs cannot be ruled out (eg. Gounelle et al., 2008).

From Figure 5 it is notable that many of the $T_j<3$ events show comparatively high heights relative to their speed compared to objects with $T_j>3$. In the interval 18 – 22 km/s half of this population falls along the very top of the $H_{max}$ distribution, suggestive of a weaker structure. Most plot in the type II TPFM region, though several are in the top (or even the middle) of the type I field, emphasizing that individual events can behave differently. The USG 20130731 event has a typical large HTC-type orbit (and $T_j \sim 1.6$) but an $H_{max}$ (29 km) and speed 18 km/s, more consistent with a strong object (and nearly identical $H_{max}$, v combination as the Buzzard Coulee meteorite fall, consistent with a fireball strength capable of producing meteorites). Interestingly, USG 20150311 has $T_j \sim$ 1.4 and nearly the same speed, entry angle and energy (and hence same mass) as USG 20130731 but an $H_{max}$ = 35 km. The significance of this difference is not clear; it could simply be due to different fragmentation behavior, but a real material/strength difference is also possible. Strong meteoroids in HTC-type orbits have been noted before (Spurný and Borovička, 1999) but such meteoroids are rare among the cm-sized population [(<0.1%) according to Gounelle et al (2008)].

Finally, three events showed peak brightness at low altitudes relative to their entry speed, suggesting stronger than typical material. USG 20091008 was an airburst over Sulawesi, Indonesia reported widely in infrasonic records (Silber et al., 2011) having a total energy above 30 kT with speed of 19 km/s and $H_{max}$ = 19.1 km. Similarly, USG 19940201 (McCord et al, 1995) was a large (30-50 kT) event at 25 km/s with an $H_{max}$ of 24 km. These $H_{max}$ are quite low, even considering their large energies (for comparison note that Chelyabinsk which is much more energetic plots in the middle of the population in Figure 5 and very near the much smaller, but physically similar Buzzard Coulee fall). For USG 19940201, entry modelling matching the lightcurve and speed as a function of height by Popova and Nemtchinov (1996) led to the conclusion that it was likely an iron meteoroid, a conclusion consistent with the event plotting near the group 0 line from TPFM simulations. However, it is also possible this is simply a stronger than typical chondritic object, the large possible range in strengths being emphasized by the exceptionally strong Carancas meteorite fall (Borovička and Spurný, 2008) which was produced by a H4-5 chondrite. The third event, USG 20141126, is much smaller (only 0.35 kT) and given the low $H_{max}$ is an even stronger candidate for a strong (possibly iron?) type object. These three objects are the strongest candidates in our sample for either true iron meteoroids (or strong, nearly monolithic chondritic bodies) showing relatively little or very late stage fragmentation.

## 4. Discussion

As originally proposed by Ceplecha (1994) in his study of meter-scale impactors, the majority of such small NEOs colliding with the Earth appeared to be weak cometary objects. As we have shown earlier, this conclusion was largely based on use of a very low luminous efficiency and did not entirely account for fragmentation. Taken together, these systematic effects increased the apparent masses, particularly for weaker bodies with high ablation coefficients, by up to an order of magnitude. This made what appeared to be large (weak) cometary bodies appear to dominate the influx at meter to tens of meter sizes. We suggest these were much smaller objects, more probably in the tens of cm category. This original inference was supported by early flux estimates from telescopic data (Rabinowitz, 1993) which suggested an unusually high flux of ~10m NEOs, close to a factor of ~100 above current estimates (eg. Boslough et al., 2015).

We can check the contention that the presumed meter-class cometary bodies were in fact smaller than reported in Ceplecha (1994) by using the Brown et al (2002) relation for the expected impact interval as a function of energy, using the assumed values of average speed, bulk density (assumed to be ordinary-chondrite-like) and mean luminous efficiency for conversion of total impact energy to estimate size as described in Brown et al (2002). In general we expect the conversion to size for cometary objects to be different than that used in Brown et al (2002), with the bulk density being lower than assumed, but the luminous efficiency being higher. As these two effects work in opposite directions we expect them to somewhat cancel out, but do not try to include explicitly the effect on sizes relative to the Brown et al (2002) assumptions and caution that there may be some systematic size effect remaining. With this in mind, we recall that the Brown et al (2002) impact frequency was based on 8 years of global data and several hundred meter-scale impacts observed by US Government sensors. We note that any one ground-based camera location has a 50% probability of detecting a meter-scale or larger impactor every 30-35 years (taking into account day-night and weather conditions for mid-northern latitude sites following the discussion in Oberst et al (1998)). From the Ceplecha (1994) analysis, a total of 13 meter-scale or larger bolides were identified as having been recorded by ground-based networks as of the early-1990s. This included one MORP event, five PN events and 7 EN fireballs. The probability of MORP detecting such a one meter-sized or larger impactor is only 15% making this somewhat unlikely. Using our estimates of the total integrated network time-area products of the camera networks ($3\times10^{11}$ km$^2$h to end of 2014 for the EN with a value roughly half this number applicable to the time of Ceplecha's 1994 study) and the global airburst rate estimated in Brown et al (2002), we estimate that the PN and EN would have had a 0.1% and <0.01% probability respectively to have recorded 5 and 7 meter-sized or larger events. However, from Table 1 & 2, we see that our revised number of meter-scale impactors for PN and EN are 1 and 3 respectively (through to 1993 appropriate to the Ceplecha (1994) dataset) and none for MORP. The probability that the PN would detect just one events is 36% and the EN two is 25%; more consistent with the global flux estimates from Brown et al (2002). Over the last ~20 years, the EN has recorded an additional meter-scale event (for a total of three),. The probability using the Brown et al (2002) impact frequency given the EN total time-area network coverage of $3\times10^{11}$ km$^2$h of detecting three or more is ~60% which is reasonable given the small number statistics. Thus we conclude that our revised smaller masses for the events used as the basis for Ceplecha's (1994) study (which we now believe to be cometary but smaller than one meter) are more consistent with modern global impact rates by bodies at these sizes.

Figure 6 shows the aerodynamic pressure at $H_{Peak}$ as a function of initial mass. There is no clear variation in the height of peak brightness (and by proxy, given our assumptions, in strength) with mass as predicted by various strength theories (eg. Holsapple, 2007) across ~5 orders of magnitude in meteoroid mass. Popova et al (2011) and Popova and Nemtchinov (2002) have also investigated the strength of large meteoroids impacting Earth's atmosphere and also find no mass dependence with strength. They suggest typical breaking strengths for meter-scale bodies lies in the range of 0.1 MPa to a few MPa, broadly consistent with our results. Some of this scatter may also be due to meteoroid shape effects. We also note that there are at least some meter-scale meteoroids with exceptionally high global strengths as evidenced by the survival of the chondritic Carancas meteorite to form an impact crater (Borovička and Spurny, 2008).

Figure 7 shows the distribution of peak heights of luminosity. We expect an upward trend in $H_{peak}$ with speed, as dynamic pressure scales as $v^2$. In fact, examination of Figure 5 seems to suggest a slight downward trend between 16 – 26 km/s, but this is not conclusive and may be small number statistics. Our median $H_{peak}$ is 31 km and the mean is 33 km. From Figure 7 it is apparent that >90% of all meter-class impactors have 20<$H_{peak}$<40 km.

The two exceptionally weak objects in our sample, Sumava and USG 20131121 have $T_j$ >3 and are likely asteroidal based solely on their orbits. Detailed analysis of Sumava (Borovička and Spurný, 1996) has suggested it is physically more cometary in nature based on its high ablation coefficient, extensive fragmentation under very low dynamic pressure and disintegration into large numbers of small particles, suggesting high microporosity. A chondritic rubble-pile, in contrast, would be more likely to fragment at high altitude into a small number of larger fragments, in addition to dust, similar to the behaviour of Almahata Sitta. It is quite likely, however, that a true "rubble-pile" assemblage consisting of many smaller (but somewhat coherent) sub-units which fragments at high altitude would not begin significant ablation until much lower, mimicking a stronger object. In this sense, high altitude ablation is perhaps most consistent with fragmentation into many smaller fragments, reflecting high microporosity, as suggested by Borovička and Spurný (1996). No further detailed data are available for USG 20131121. Its exceptionally high altitude of peak brightness make it the best candidate among our events for a very weak object.

The overall trends in Figure 5 suggest that there is a bit more than an order of magnitude spread in the strength of meter-scale objects impacting Earth and that this population is similar to the well documented meteorite producing large fireballs. A minority of objects (~10%-20%) are somewhat weaker than the overall population, but even these may be best viewed as part of a continuum, which extends from very strong objects whose peak luminosity occurs at heights which suggests almost no fragmentation (ie. consistent with single body ablation such as USG 20091008) to those just barely into the type IIIa fireball category. Only two objects (Sumava and USG 20131121) stand out as unquestionably weaker than the rest of the impacting population.

## 5. Conclusions

We have examined a suite of 59 meter-scale Earth impactors having a median energy of 0.4 kT with known orbits, speeds and heights of peak luminosity. The majority of our events were collected by US Government sensors and have little ancillary information. However, we find that calibrated events from ground-based fireball networks and meteorite-producing fireballs allow for a basic framework to examine the population of USG events, because they follow the trends in peak brightness vs speed that are expected from ablation modelling.

Our analysis suggests that use of the height of peak brightness as a criterion for the strength for larger fireballs is reasonable. This is supported by fireball events for which detailed PE data are available, and which fall into the expected TPFM (Triggered progressive fragmentation model) bands and/or agree with the fireball events producing recovered meteorites.

Within the limitations of our still small sample statistics our major conclusions include:

1. A lower limit of ~10% of our population have orbital characteristics and relatively weak strengths potentially consistent with cometary material. Orbital characteristics alone suggest a JFC fraction of 5-10% and a similar HTC (Halley-type comets) fraction. These values are similar to those found from studies examining the dynamical and spectroscopic properties of larger NEAs (eg. DeMeo and Binzel (2004))

2. Most of our population have orbits and physical characteristics comparable to recovered stony/chondritic bodies delivered from the inner main-belt.

3. The similarity in the height of peak brightness as a function of speed behavior of most of our events compared to well documented meteorite-producing fireballs suggests that the majority of meter-scale NEOs are globally weak with strengths of order 0.1 – 1 MPa. This is consistent with the conclusions of Popova et al (2011). We do not find that we can distinguish different materials (ie. ordinary chondrite vs. carbonaceous chondrite) using peak height alone, but only broad global strength properties and that the latter is similar for many different meteorite-types.

4. The Taurid shower and its sub-components produce the only significant shower/NEA associations among our dataset; no major meteor showers show similar orbits to any of our meter-scale events.

5. Two events in our data have exceptionally weak structure; Sumava and USG 20131121. The former is well documented and almost certainly cometary-like while the latter is our best candidate for a truly weak NEA compared to the population as a whole.

6. Three events show evidence for being unusually strong. The USG 20091008 (Sulawesi, Indonesia) airburst, in particular, is the only object among our data displaying a peak brightness below 20 km altitude, suggestive of a physically strong body.

7. Our overall aei orbital distributions are a good match to NEA model predictions of the impactor population at the Earth (eg. Veres et al, 2009). This suggests that the majority of meter-scale NEOs are in fact derived from the main belt. We find no retrograde orbits among our meter-scale impactors.

8. There is a wide (more than order of magnitude) spread in strengths of meter-scale objects, but no clear trend in strength with size/mass.

9. An earlier suggestion by Ceplecha (1994) that the majority of meter-tens of meter-scale NEOs are weak, cometary-like objects is not supported by the current study. We suggest that application of very low values for luminous efficiency and incomplete treatment of fragmentation produced overestimated masses, a conclusion also reached by Ceplecha and ReVelle (2005) in development of their FM (Fragmentation model) as applied to fireballs.

**Acknowledgements**

PGB thanks the Canada Research Chair program. This work is supported in part by the Natural Sciences and Engineering Research Council of Canada and the NASA Meteoroid Environment Office through co-operative agreements NN61AB76A and NN65AC94A. Helpful discussions with O. Popova and P Spurný are gratefully acknowledged. Helpful reviews by J. Borovicka, M. Boslough and an anonymous reviewer greatly improved the manuscript.

Table 1. Meter-scale impactors detected as fireballs by ground-based networks. The orbital elements for each event are given including semi-major axis (a), eccentricity (e), orbital inclination (inc), longitude of the ascending node (Ω), argument of perihelion (ω), perihelion (q), aphelion (Q), initial kinetic energy in kilotons (kT) of TNT equivalent where 1 kT = 4.184×10$^{12}$ J, mass of initial body (kg), diameter of initial body (m), speed at the top of the atmosphere, height of maximum brightness, $Z_r$ - local radiant zenith angle (degs), and fireball type following the Ceplecha and McCrosky (1976) classification. The original literature reference for the orbit/speed/mass/energy/fireball type data is also indicated. Note that the diameter is computed assuming a bulk density appropriate to the fireball type following Ceplecha et al (1998) (see caption in figure 5 for specific values of density used for each fireball class).

| Event | a | e | inc | Ω | ω | q | Q | Tj | Energy | Mass | D | Speed | Height | Zr | Type | Reference |
|---|---|---|---|---|---|---|---|---|---|---|---|---|---|---|---|---|
| | (AU) | | degs | degs | degs | (AU) | (AU) | | (kT) | (kg) | (m) | (km/s) | (km) | (degs) | | |
| 19661211 (PN 39470) | 2.11 | 0.710 | 0.8 | 259.00 | 265.1 | 0.613 | 3.61 | 3.36 | 0.12 | 1.74E+03 | 1.0 | 23.6 | 33.0 | 56.0 | I | 1,2 |
| 19741204 (Sumava) (EN041274) | 1.90 | 0.755 | 2.2 | 252.65 | 283.0 | 0.466 | 3.34 | 3.53 | 0.43 | 5.00E+03 | 3.2 | 26.9 | 67 | 62.6 | IIIb | 3 |
| 20011117 (EN171101) | 1.33 | 0.484 | 7.4 | 235.39 | 266.8 | 0.684 | 1.97 | 4.80 | 0.18 | 4.30E+03 | 1.3 | 18.5 | 25.0 | 50.0 | I | 4,5,6 |
| | | | | | | | | | | | | | | | | |
| | | | | | | | | | | | | | | | | |

References are 1 – Ceplecha (1994); 2 – Popova (1996); 3 – Borovička and Spurný (1996); 4- Shrbeny (2009); 5 – Spurný and Porubcan (2002); 6 – Svoren et al (2008)

Table 2. Meter-sized impactors detected as fireballs where meteorites were also recovered. The meteorite fall name is given together with the letter code used in Figure 5 . The orbital elements (J2000.0) for each event are given including semi-major axis (a), eccentricity (e), orbital inclination (inc), longitude of the ascending node ($\Omega$)**,** argument of perihelion ($\omega$), perihelion (q), aphelion (Q), initial kinetic energy in kilotons (kT) of TNT equivalent where 1 kT = 4.184×10$^{12}$ J, mass of initial body (kg), diameter of initial body (m), speed at the top of the atmosphere, height of maximum brightness, luminous end height (used for calculation of the fireball type), local radiant zenith angle (degs), fireball type following the Ceplecha and McCrosky (1976) classification and original literature reference for the orbit/speed/mass/energy/fireball type data. Met denotes the meteorite type recovered. Mx represents mixed meteorite types which were recovered, specifically of LL3.5, H5, primitive achondrite for Benesov and anomalous Ureilite, EL,EH,H,L,LL,CB and R for Almahata Sitta.  Note that the diameter is computed using the bulk density of the recovered meteorites.  For Almahata Sitta, Buzzard Coulee, Chelyabinsk and Kosice the independent estimates from US Government sensors are shown in bold for comparison.

| Event | a | e | inc | $\Omega$ | $\omega$ | q | Q | Tj | En | Mass | D | Speed | Height | End Height | Zr | Type | Met | Reference |
|---|---|---|---|---|---|---|---|---|---|---|---|---|---|---|---|---|---|---|
| | (AU) | | degs | degs | degs | (AU) | (AU) | | (kT) | (kg) | (m) | (km/s) | (km) | (km) | (degs) | | | |
| Benesov (B) | 2.48 | 0.627 | 24.0 | 47.00 | 218.4 | 0.925 | 4.04 | 3.08 | 0.20 | 3.5E+03 | 1.3 | 21.1 | 24.3 | 19.2 | 9.5 | I | Mx | 1,2 |
| Almahata Sitta (AS)[2008 TC3] | 1.31 | 0.310 | 2.5 | 194.10 | 234.5 | 0.904 | 1.72 | 4.93 | 0.92 | 5.00E+04 | 3.1 | 12.4 | 37.0 | 32.7 | 70.0 | IIIa | Mx | 3,4 |
| | **1.64** | **0.43** | **3.9** | **14.1** | **36.3** | **0.94** | **2.35** | **4.2** | **1** | **4.7E+04** | **3.0** | **13.3** | **38.9** | | **84** | | | |
| Buzzard Coulee (BC) | 1.23 | 0.220 | 25.5 | 238.90 | 212.0 | 0.959 | 1.50 | 5.09 | 0.31 | 8.00E+03 | 1.7 | 18.0 | 31.0 | 13 | 23.0 | I | H4 | 5,6,7 |
| | **0.79** | **0.26** | **10.3** | **238.93** | **3.2** | **0.58** | **0.99** | **7.4** | **0.41** | **2.06E+04** | **2.3** | **12.9** | **28.2** | | **53.0** | | | |
| Chelyabinsk (C) | 1.72 | 0.571 | 5.0 | 326.46 | 107.7 | 0.738 | 2.7 | 3.97 | 500 | 1.2E+07 | 19 | 19.0 | 30.0 | 13 | 72 | II | LL5 | 8,9,10 |
| | **1.71** | **0.56** | **4.1** | **326.41** | **109.7** | **0.75** | **2.67** | **3.99** | **440** | **1.1E+07** | **18** | **18.6** | **23.0** | | **74** | | | |
| Kosice (K) | 2.71 | 0.647 | 2.0 | 340.07 | 204.2 | 0.957 | 4.46 | 3.02 | 0.13 | 3.50E+03 | 1.2 | 15.0 | 36.0 | 17.4 | 30.0 | II | H5 | 11,12 |
| | **2.70** | **0.65** | **3.2** | **340.08** | **204** | **0.957** | **4.44** | **3.03** | **0.44** | **1.6E+04** | **2.0** | **15.1** | **37.0** | | **27.1** | | | |
| Maribo (M) | 2.48 | 0.807 | 0.1 | 297.12 | 279.2 | 0.479 | 4.48 | 2.91 | 0.14 | 1.50E+03 | 1.1 | 28.3 | 37.1 | 30.5 | 60.0 | II | CM2 | 13 |
| Park Forest (PF) | 2.53 | 0.680 | 3.2 | 6.10 | 237.5 | 0.810 | 4.25 | 3.08 | 0.32 | 7.00E+03 | 1.6 | 19.5 | 29.0 | 18 | 29.0 | I/II | L5 | 14,15,16 |
| Peekskill (P) | 1.49 | 0.410 | 4.9 | 17.00 | 307.6 | 0.879 | 2.10 | 4.47 | 0.13 | 5.00E+03 | 1.4 | 14.7 | | | 87.0 | | H6 | 17,18 |
| Sutter's Mill (SM) | 2.59 | 0.824 | 2.4 | 32.77 | 77.8 | 0.456 | 4.7 | 2.81 | 3.91 | 4.00E+04 | 3.3 | 28.6 | 47.6 | 30 | 64.0 | II/IIIa | CM2 | 19 |
| Tagish Lake (TL) | 1.98 | 0.550 | 2.0 | 297.90 | 224.4 | 0.891 | 3.07 | 3.66 | 1.79 | 6.00E+04 | 4.2 | 15.8 | 32.0 | 29 | 72.0 | II/IIIa | C2 | 20,21 |

References are 1 –Spurný et al (2014) ; 2 - Borovička and Spurný (1996); 3- Jenniskens et al (2009); 4 – Kohout et al (2011); 5 – Brown et al (2011); 6 – Hildebrand et al (2009); 7 – Milley et al (2010); 8 - Borovička et al (2013); 9 - Popova et al (2013); 10 – Brown et al (2013b); 11 – Borovička et al (2013b); 12 – Kohout et al (2014); 13 – Spurný et al (2013); 14 – Brown et al (2004); 15 - Macke (2010); 16 – Simon et al (2004); 17 – Brown et al (1994); 18 – Borovička and Kalenda (2003); 19 – Jenniskens et al (2012); 20 – Hildebrand et al (2006); 21 – Brown et al (2002b)

Table 3. Meter-sized impactors detected as fireballs by US Government sensors (USG). The orbital elements (J2000.0) for each event are given including semi-major axis (a), eccentricity (e), orbital inclination (inc), longitude of the ascending node (Ω)**,** argument of perihelion (ω), perihelion (AU), aphelion (Q), initial kinetic energy in kilotons (kT) of TNT equivalent where 1 kT = 4.184x10$^{12}$ J, mass of initial body (kg), diameter of initial body (m), speed at the top of the atmosphere, height of maximum brightness, local radiant zenith angle (degs), Note that the diameter is computed using a fixed bulk density of 3700 kgm$^{-3}$. Where a reference is not given, data are directly extracted from the JPL Fireball webpage (http://neo.jpl.nasa.gov/fireballs/).

| Event | a | e | inc | Ω | ω | q | Q | Tj | En | Mass | D | Speed | Height | Zr | Reference |
|---|---|---|---|---|---|---|---|---|---|---|---|---|---|---|---|
| | (AU) | | degs | degs | degs | (AU) | (AU) | | (kT) | (kg) | (m) | (km/s) | (km) | (degs) | |
| 19940201-223809 | 2.10 | 0.740 | 2.0 | 132.92 | 268.0 | 0.546 | 3.65 | 3.33 | 30.30 | 4.06E+05 | 6.17 | 25.0 | 24.0 | 45.5 | 1,2 |
| 19990114-080605 | 1.90 | 0.490 | 14.0 | 114.00 | 353.0 | 0.969 | 2.83 | 3.76 | 9.80 | 3.65E+05 | 5.95 | 15.0 | 35.0 | 54.6 | 3 |
| 20040903-120721 | 0.86 | 0.178 | 12.2 | 341.26 | 196.9 | 0.710 | 1.02 | 6.81 | 13.00 | 6.44E+05 | 7.20 | 13.0 | 25.0 | 48.0 | 4 |
| 20091008-025700 | 1.20 | 0.548 | 14.1 | 194.84 | 72.2 | 0.541 | 1.85 | 5.13 | 33.00 | 5.70E+05 | 6.91 | 22.0 | 19.1 | 22.5 | |
| 20091121-205300 | 0.84 | 0.588 | 56.4 | 239.55 | 319.3 | 0.346 | 1.33 | 6.57 | 18.00 | 1.47E+05 | 4.40 | 32.0 | 38.0 | 81.3 | |
| 20101225-232400 | 1.01 | 0.394 | 16.4 | 273.93 | 69.7 | 0.611 | 1.40 | 5.94 | 33.00 | 8.12E+05 | 7.77 | 18.5 | 26.0 | 29.1 | |
| 20120826-145547 | 1.24 | 0.254 | 3.4 | 153.75 | 233.9 | 0.926 | 1.56 | 5.13 | 0.68 | 3.53E+04 | 2.73 | 12.7 | 36.0 | 9.4 | |
| 20120827-065743 | 1.68 | 0.747 | 20.0 | 154.24 | 69.1 | 0.424 | 2.93 | 3.81 | 0.22 | 2.21E+03 | 1.08 | 28.9 | 38.7 | 48.9 | |
| 20120910-010332 | 0.93 | 0.282 | 20.2 | 347.58 | 121.3 | 0.671 | 1.20 | 6.34 | 0.08 | 2.40E+03 | 1.12 | 16.9 | 23.8 | 59.1 | |
| 20120918-193439 | 1.37 | 0.383 | 19.6 | 176.12 | 114.9 | 0.844 | 1.89 | 4.70 | 0.67 | 1.68E+04 | 2.13 | 18.3 | 28.1 | 46.7 | |
| 20121002-163838 | 0.89 | 0.354 | 3.0 | 9.76 | 230.2 | 0.572 | 1.20 | 6.65 | 1.20 | 4.24E+04 | 2.91 | 15.4 | 35.0 | 17.1 | |
| 20121009-005455 | 2.17 | 0.540 | 4.5 | 15.99 | 356.0 | 0.999 | 3.34 | 3.48 | 0.58 | 2.66E+04 | 2.49 | 13.5 | 27.8 | 75.8 | |
| 20121019-005455 | 1.53 | 0.349 | 9.2 | 26.62 | 6.1 | 0.995 | 2.06 | 4.41 | 0.08 | 3.94E+03 | 1.32 | 13.2 | 29.3 | 6.2 | |
| 20121120-203731 | 1.79 | 0.468 | 8.6 | 58.81 | 28.6 | 0.951 | 2.62 | 3.94 | 0.09 | 3.64E+03 | 1.28 | 14.3 | 33.3 | 25.0 | |
| 20130421-062312 | 1.76 | 0.476 | 1.4 | 60.02 | 197.3 | 0.922 | 2.60 | 3.98 | 2.50 | 9.06E+04 | 3.74 | 15.2 | 40.7 | 49.2 | |
| 20130430-084038 | 1.07 | 0.122 | 7.2 | 55.96 | 251.4 | 0.936 | 1.20 | 5.77 | 10.00 | 5.54E+05 | 6.84 | 12.3 | 21.2 | 50.5 | |
| 20130727-083036 | 1.60 | 0.575 | 17.1 | 124.35 | 264.5 | 0.680 | 2.52 | 4.12 | 0.36 | 6.17E+03 | 1.53 | 22.1 | 26.5 | 67.0 | |
| 20130730-023658 | 2.37 | 0.632 | 9.7 | 307.07 | 308.6 | 0.872 | 3.87 | 3.23 | 1.00 | 2.37E+04 | 2.40 | 18.8 | 25.6 | 59.8 | |
| 20130731-035014 | 11.70 | 0.916 | 1.8 | 127.94 | 156.0 | 0.983 | 22.42 | 1.65 | 0.22 | 5.78E+03 | 1.50 | 17.9 | 29.1 | 46.5 | |

| Event | a | e | inc | Ω | ω | q | Q | Tj | En | Mass | D | Speed | Height | Zr | Reference |
|---|---|---|---|---|---|---|---|---|---|---|---|---|---|---|---|
| | (AU) | | degs | degs | degs | (AU) | (AU) | | (kT) | (kg) | (m) | (km/s) | (km) | (degs) | |
| 20131012-160645 | 0.96 | 0.144 | 9.4 | 199.41 | 64.3 | 0.817 | 1.09 | 6.29 | 3.50 | 1.74E+05 | 4.65 | 13.0 | 22.2 | 49.1 | |
| 20131121-015035 | 0.77 | 0.294 | 2.0 | 59.00 | 185.0 | 0.544 | 1.00 | 7.50 | 0.23 | 1.26E+04 | 1.94 | 12.4 | 59.3 | 41.2 | |
| 20131208-031009 | 0.98 | 0.079 | 5.6 | 256.50 | 279.8 | 0.901 | 1.06 | 6.18 | 0.20 | 1.21E+04 | 1.91 | 11.8 | 23.5 | 43.7 | |
| 20131223-083057 | 0.85 | 0.374 | 4.9 | 271.52 | 313.2 | 0.533 | 1.17 | 6.86 | 0.43 | 1.52E+04 | 2.06 | 15.4 | 34.3 | 63.1 | |
| 20140112-160048 | 3.39 | 0.716 | 9.1 | 112.21 | 19.9 | 0.960 | 5.81 | 2.65 | 0.24 | 7.75E+03 | 1.65 | 16.1 | 37.0 | 13.8 | |
| 20140329-134541 | 1.90 | 0.534 | 3.9 | 188.63 | 47.3 | 0.887 | 2.92 | 3.76 | 0.13 | 4.10E+03 | 1.33 | 16.3 | 30.7 | 21.9 | |
| 20140508-194237 | 0.93 | 0.498 | 5.9 | 227.94 | 127.6 | 0.469 | 1.40 | 6.30 | 2.40 | 5.57E+04 | 3.18 | 19.0 | 35.4 | 6.6 | |
| 20140516-124248 | 0.82 | 0.447 | 1.6 | 55.41 | 321.4 | 0.454 | 1.19 | 7.04 | 0.82 | 2.55E+04 | 2.45 | 16.4 | 44.0 | 23.8 | |
| 20140516-200628 | 2.26 | 0.618 | 12.6 | 55.68 | 127.7 | 0.862 | 3.65 | 3.32 | 0.40 | 9.68E+03 | 1.78 | 18.6 | 30.8 | 82.0 | |
| 20140626-055441 | 1.07 | 0.049 | 0.3 | 108.67 | 213.7 | 1.014 | 1.12 | 5.78 | 0.20 | 1.34E+04 | 1.98 | 11.2 | 28.5 | 50.9 | |
| 20140628-024007 | 2.89 | 0.660 | 9.3 | 96.19 | 155.8 | 0.982 | 4.80 | 2.91 | 0.67 | 2.11E+04 | 2.30 | 16.3 | 26.3 | 51.4 | |
| 20140823-062941 | 1.35 | 0.336 | 20.7 | 329.87 | 57.3 | 0.894 | 1.80 | 4.76 | 7.60 | 2.05E+05 | 4.92 | 17.6 | 22.2 | 47.9 | |
| 20141014-102503 | 2.66 | 0.657 | 7.0 | 200.81 | 218.3 | 0.912 | 4.41 | 3.03 | 0.10 | 2.93E+03 | 1.19 | 16.9 | 27.2 | 38.5 | |
| 20141104-201330 | 1.18 | 0.406 | 3.2 | 222.10 | 271.0 | 0.701 | 1.66 | 5.28 | 0.45 | 1.46E+04 | 2.04 | 16.1 | 22.2 | 47.1 | |
| 20141126-174016 | 2.90 | 0.726 | 0.2 | 64.22 | 302.8 | 0.794 | 5.00 | 2.82 | 0.32 | 6.77E+03 | 1.58 | 19.9 | 37.0 | 47.2 | |
| 20141126-231651 | 2.27 | 0.745 | 6.8 | 64.48 | 271.8 | 0.579 | 3.96 | 3.17 | 0.35 | 4.58E+03 | 1.38 | 25.3 | 23.3 | 37.6 | |
| 20141128-114718 | 1.29 | 0.257 | 10.2 | 66.06 | 328.8 | 0.958 | 1.62 | 4.98 | 1.70 | 7.93E+04 | 3.58 | 13.4 | 26.1 | 43.6 | |
| 20141212-064811 | 1.43 | 0.311 | 0.1 | 254.64 | 177.3 | 0.985 | 1.87 | 4.64 | 0.11 | 6.35E+03 | 1.54 | 12.0 | 26.3 | 30.2 | |
| 20141213-025352 | 1.70 | 0.557 | 20.8 | 260.81 | 109.8 | 0.752 | 2.64 | 3.96 | 0.15 | 2.67E+03 | 1.16 | 21.7 | 30.7 | 67.3 | |
| 20150102-133919 | 9.38 | 0.900 | 8.0 | 281.60 | 207.4 | 0.938 | 17.82 | 1.71 | 0.07 | 1.87E+03 | 1.03 | 18.1 | 38.1 | 71.6 | |
| 20150107-010559 | 4.78 | 0.930 | 20.7 | 106.19 | 112.0 | 0.335 | 9.23 | 1.75 | 0.40 | 2.63E+03 | 1.15 | 35.7 | 45.5 | 46.4 | |
| 20150109-104111 | 1.40 | 0.449 | 10.9 | 108.65 | 288.0 | 0.771 | 2.03 | 4.63 | 0.41 | 1.12E+04 | 1.87 | 17.5 | 36.0 | 43.9 | |
| 20150226-220624 | 2.04 | 0.514 | 28.9 | 337.78 | 177.6 | 0.990 | 3.08 | 3.49 | 0.53 | 9.97E+03 | 1.79 | 21.1 | 33.7 | 17.8 | |
| 20150304-043005 | 1.61 | 0.523 | 6.4 | 343.08 | 109.6 | 0.768 | 2.45 | 4.17 | 0.18 | 4.65E+03 | 1.39 | 18.0 | 39.8 | 36.5 | |
| 20150311-061859 | 18.90 | 0.947 | 16.6 | 350.14 | 159.1 | 1.002 | 36.80 | 1.45 | 0.23 | 4.86E+03 | 1.41 | 19.9 | 35.2 | 45.5 | |
| 20150330-213352 | 1.52 | 0.385 | 0.7 | 189.86 | 39.2 | 0.935 | 2.11 | 4.42 | 0.20 | 8.79E+03 | 1.72 | 13.8 | 33.1 | 34.8 | |
| 20150408-040631 | 2.16 | 0.630 | 8.2 | 17.82 | 118.8 | 0.801 | 3.53 | 3.40 | 0.49 | 1.11E+04 | 1.86 | 19.2 | 36.3 | 80.7 | |

References are 1 –McCord et al (1995); 2 – Tagliaferri et al (1995); 3 – Pack et al (1999); 4 – Klekociuk et al (2005)

Table 4. Fireballs detected by US Government sensors with reported heights of peak luminosity ($H_{USG}$) also having independent estimates of the height of peak brightness ($H_{Peak}$) as described in each of the associated references.

| Event (Date) | $H_{USG}$ (km) | $H_{Peak}$ (km) | Reference |
|---|---|---|---|
| Chelyabinsk (2013-02-15) | 23 | 30 | 1,2 |
| **Kosice (2010-02-28)** | 37 | 36 | 3 |
| Buzzard Coulee (2008-11-21) | 28.2 | 31 | 4 |
| Almahatta Sitta (2008-10-07) | 37 | 36 | 5 |
| Moravka (2000-05-06) | 36 | 33 | 6 |
| Tagish Lake (2000-01-18) | 35 | 32 | 7,8 |
| El Paso fireball (1997-10-09) | 27 | 29 | 9 |

1 - Borovička et al (2013); 2 - Popova et al (2013); 3 – Borovička et al (2013b); 4- Milley (2010); 5 - Borovička and Charvat (2009); 6 - Borovička et al (2003); 7 - Brown et al (2002b); 8-Hildebrand et al (2006); 9 - Hildebrand et al (1999)

Table 5. Meter-sized impactors and possible NEA associations based on the orbital similarity Drummond D' criterion. Here n=number of Monte Carlo draws for a better match, N = number of NEAs of this size or larger, 1/p is 1/coincidence probability. The color of the rows distinguishes associations: the last two entries are fireballs with 2 NEAs with less than 1:100 coincidence probabilities. The orbital elements for each object are shown, together with the Diameter (D) of each object and type (fireball type if known or asteroid spectral taxonomic type if known). Where no reference is given the data are from this paper and/or the JPL Fireball website.

| Event | D' | n | N | 1/p | a (AU) | e | inc degs | Ω degs | ω degs | q (AU) | Q (AU) | Tj | D (m) | Type | Reference |
|---|---|---|---|---|---|---|---|---|---|---|---|---|---|---|---|
| 19661211 (PN 39470) | | | | | 2.11 | 0.710 | 0.8 | 259.00 | 265.1 | 0.613 | 3.61 | 3.36 | 1.0 | I | 1,2 |
| 2201 Oljato | 0.043 | 3.9e4 | 155 | 250 | 2.17 | 0.713 | 2.5 | 75.0 | 98.2 | 0.62 | 3.72 | 3.30 | 4000 | E;Sq | 3 |
| Chelyabinsk | | | | | 1.72 | 0.571 | 5.0 | 326.5 | 107.7 | 0.74 | 2.70 | 3.99 | | I/II | 4 |
| 86039 (1999 NC43) | 0.018 | 3e6 | 227 | 13,000 | 1.76 | 0.579 | 7.1 | 311.8 | 120.6 | 0.74 | 2.78 | 3.90 | 2900 | Q | 5 |
| USG 20121120 | | | | | 1.79 | 0.468 | 8.6 | 58.8 | 28.6 | 0.95 | 2.62 | 3.95 | | | |
| 163014 (2001 UA5) | 0.033 | 6.9e4 | 608 | 110 | 1.79 | 0.446 | 10.0 | 58.7 | 27.5 | 0.99 | 2.58 | 3.96 | 1400 | Sq | 3 |
| Maribo | | | | | 2.48 | 0.807 | 0.1 | 297.1 | 279.2 | 0.48 | 4.47 | 2.91 | | II | 6 |
| 85182 (1991 AQ) | 0.044 | 8.4e4 | 479 | 175 | 2.22 | 0.776 | 0.5 | 339.7 | 242.9 | 0.50 | 3.94 | 3.16 | 1700 | Sq; | 7 |
| USG 20130430 | | | | | 1.07 | 0.122 | 7.3 | 56.0 | 251.4 | 0.94 | 1.20 | 5.78 | | | |
| 207945 (1991 JW) | 0.038 | 6.3e5 | 1760 | 360 | 1.04 | 0.119 | 8.7 | 53.9 | 301.9 | 0.92 | 1.25 | 5.89 | 640 | Q | 8 |
| USG 20140516 | | | | | 0.82 | 0.447 | 1.6 | 55.4 | 321.4 | 0.45 | 1.19 | 7.04 | | | |
| 2000 EM26 | 0.045 | 9.7e5 | 8160 | 120 | 0.82 | 0.470 | 3.9 | 345.0 | 24.2 | 0.43 | 1.20 | 7.07 | 200 | | |
| USG 20140626 | | | | | 1.07 | 0.049 | 0.3 | 108.7 | 213.7 | 1.01 | 1.12 | 5.78 | | | |
| 2014 UX7 | 0.045 | 1.0e7 | 8.8e4 | 115 | 1.11 | 0.050 | 7.3 | 208.4 | 136.4 | 1.05 | 1.16 | 5.63 | 30 | | |
| 163000 (2001 SW169) | 0.083 | 1.9e5 | 1500 | 130 | 1.25 | 0.052 | 3.6 | 8.5 | 284.8 | 1.18 | 1.31 | 5.14 | 700 | Sq | 8 |
| USG 20141212 | | | | | 1.43 | 0.311 | 0.1 | 254.6 | 177.3 | 0.99 | 1.87 | 4.64 | | | |
| 2010 WW8 | 0.016 | 8.3e7 | 1.9e5 | 430 | 1.47 | 0.310 | 1.7 | 252.4 | 177.5 | 1.01 | 1.92 | 4.56 | 20 | | |
| 2008 VA15 | 0.019 | 3.6e7 | 2.4e4 | 1500 | 1.45 | 0.305 | 1.8 | 335.4 | 96.5 | 1.01 | 1.89 | 4.59 | 90 | | |

1 – Ceplecha (1994); 2 – Popova (1996); 3 - Binzel et al (2004); 4 - Borovička et al (2013); 5 - Reddy et al (2015); 6 - Spurný et al (2013); 7 – Somers et al (2010); 8 – DeMeo et al (2014)

Table 6. The source regions for NEOs / impactors based on residence-time probability models of delivery from the main-belt/JFCs following the Bottke et al (2002) model.

| Source Region | This study | Binzel et al (2004) | Bottke et al (2002) | Greenstreet et al (2012) |
|---|---|---|---|---|
| JFC | 4% | 2% | 6% | 5% |
| Outer Main-Belt | 8% | 6% | 8% | 8% |
| 3:1 MMR | 14% | 19% | 23% | 21% |
| IMC | 26% | 27% | 25% | 25% |
| $\nu_6$ | 48% | 46% | 37% | 41% |

**Figure Captions**

Figure 1. The cumulative number of known NEAs with H< 20 matching the orbit of meter-scale fireball events with a value of D' greater than the values shown on the abscissa.

Figure 2. The observed fractional distribution of meter-sized impactor orbits in terms of semi-major axis (top), eccentricity (middle) and inclination (bottom).

Figure 3. The total normalized source region probability for all non-HTC impactor orbits in our dataset. The intermediate source regions follow the definitions given in the Bottke et al (2002) model. In this model, the immediate source regions include the 3:1 mean motion resonance with Jupiter (where a ~ 2.5 AU), the $\nu_6$ secular resonance which operates in the inner main-belt, the intermediate Mars crossers (IMC) which are the subset of Mars-crossing bodies which are also near the main belt, the Outer Belt (OB) where asteroids with a > 2.8 AU feed into various resonances and finally Jupiter-family comets (JFCs) where $2<T_j<3$.

Figure 4. Impact speed at the top of the atmosphere for all events in our dataset using 3 km/s binning.

Figure 5. The measured height of peak brightness as a function of initial in-atmosphere speed (or speed at height of peak brightness if known) for all meter-class impactors where these data are available. The triangles represent the three camera network events (summarized in Table 1). Circles are US Government sensor detected events. Meteorite

producing fireballs are shown as squares labelled according to the letter designation shown in Table 2. Also shown are contours of equivalent ram pressure as a function of height and speed (dotted black lines). The height intervals corresponding to the expected range where peak brightness is expected for a 0.4 kT energy impactor as estimated from the TPFM model of ReVelle (2005) are shown as the solid blue lines. Following the fireball type classification of CM76, the following parameters are used in the TPFM model for strength, bulk density and ablation energy for each fireball type shown:

type 0, 100 MPa, 7800 kgm$^{-3}$, 8x10$^6$ J/kg

type I, 0.7 MPa, 3700 kgm$^{-3}$, 8.5x10$^6$ J/kg

type II, 0.2 MPa, 2100 kgm$^{-3}$, 8.5x10$^6$ J/kg

type IIIa, 0.01 MPa, 750 kgm$^{-3}$, 8x10$^6$ J/kg

type IIIb, 0.001 MPa, 270 kgm$^{-3}$, 4x10$^6$ J/kg

Orbits with $T_j <2$ (which we associate with HTC/NICs) are shown in green and those having orbits with $2<T_j<3$ (JFCs) are shown are in red..

Figure 6. The aerodynamic pressure at the height of maximum luminosity ($H_{Peak}$ Presssure) as a function of mass for all meter-scale events.

Figure 7. The distribution of heights at maximum luminosity for all meter-scale airbursts in 5 km height bins.

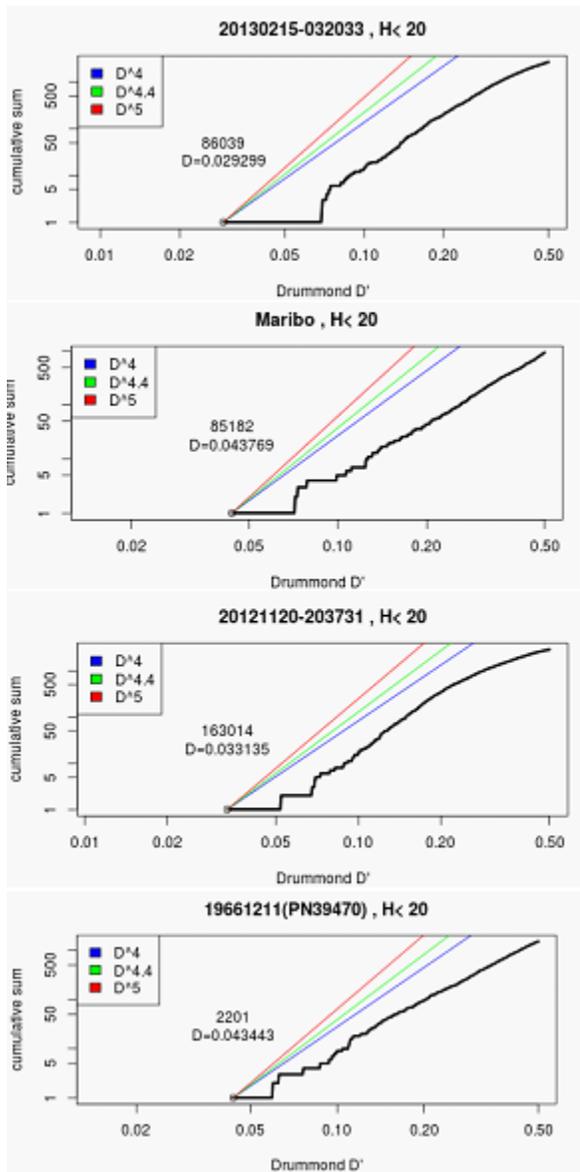

Figure 1.

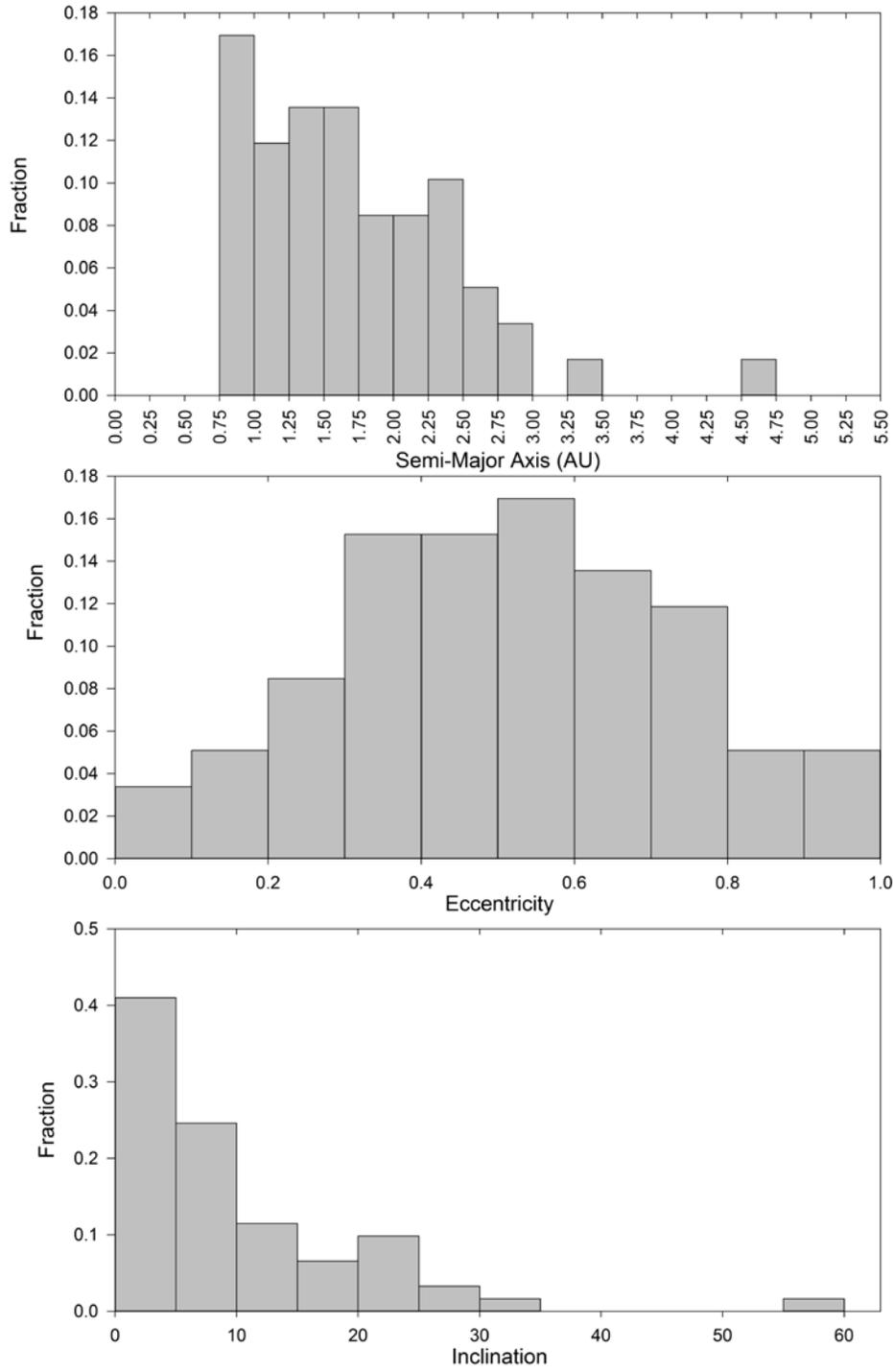

Figure 2.

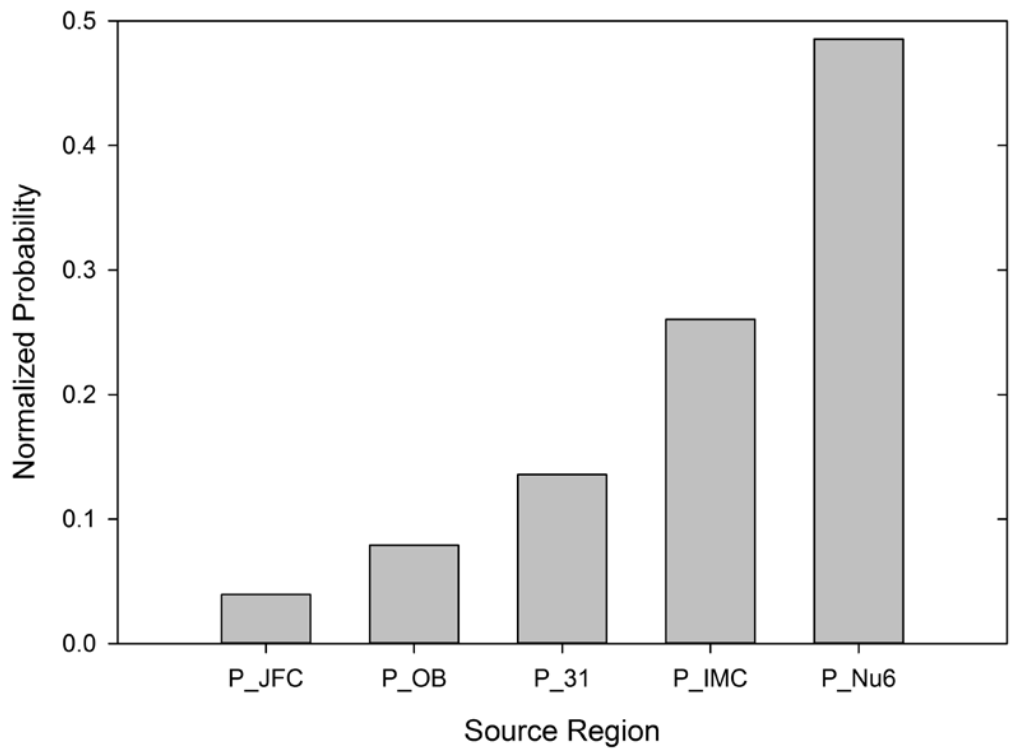

Figure 3.

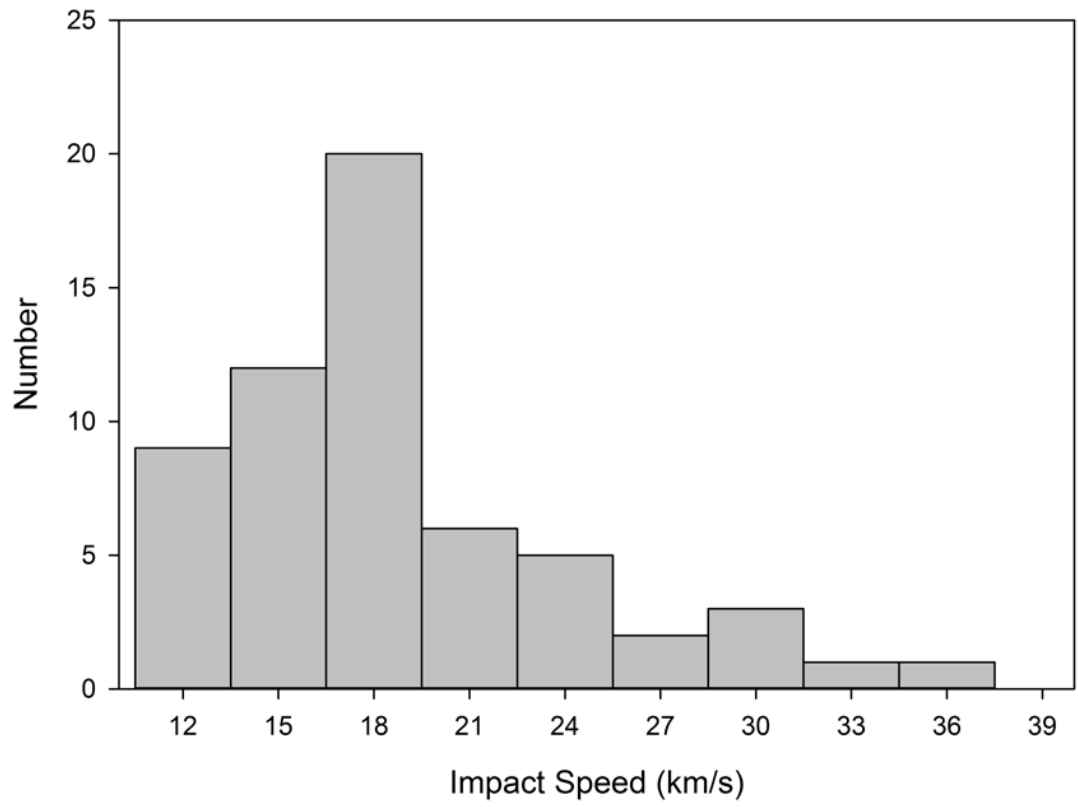

Figure 4.

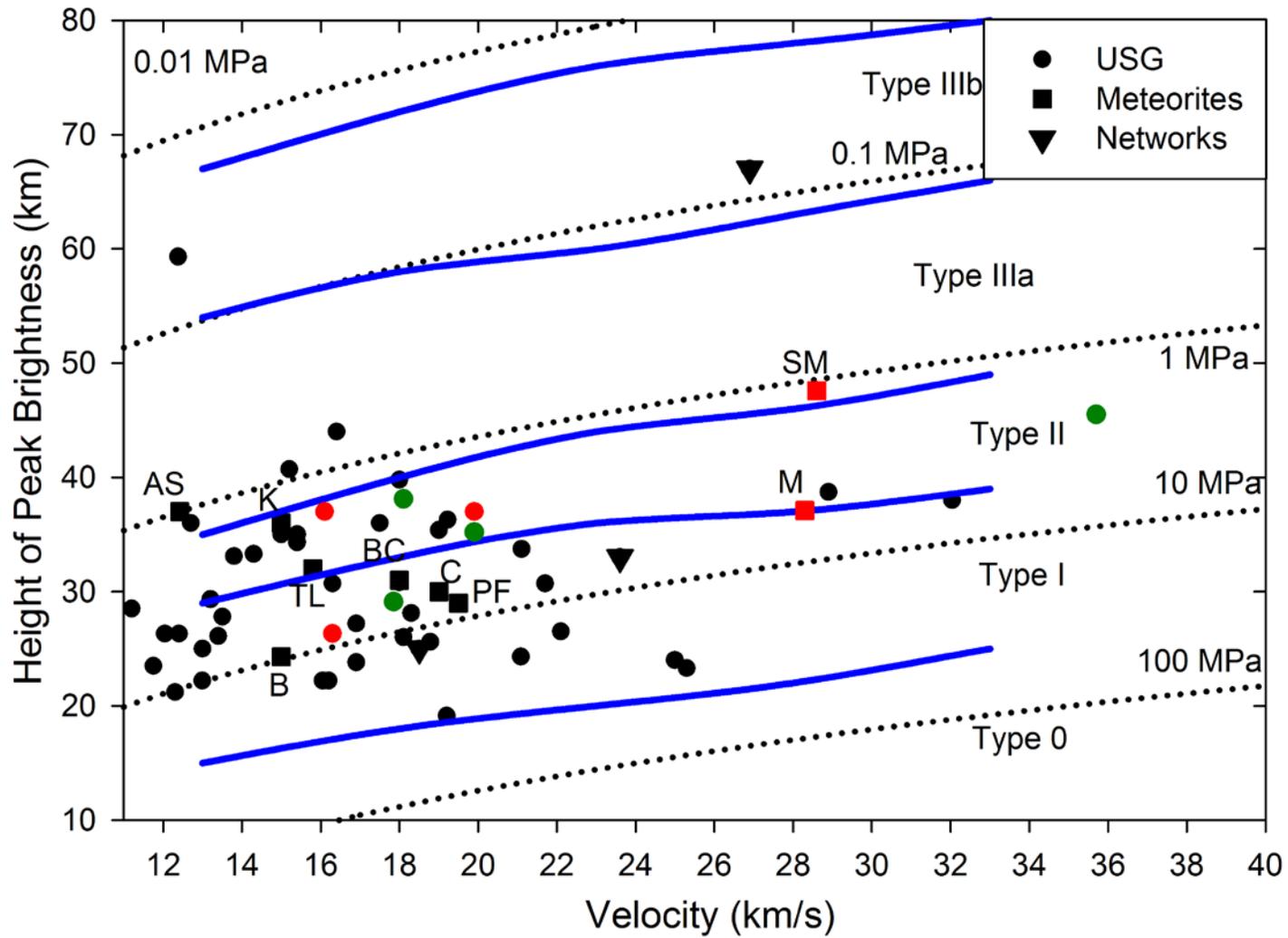

Figure 5.

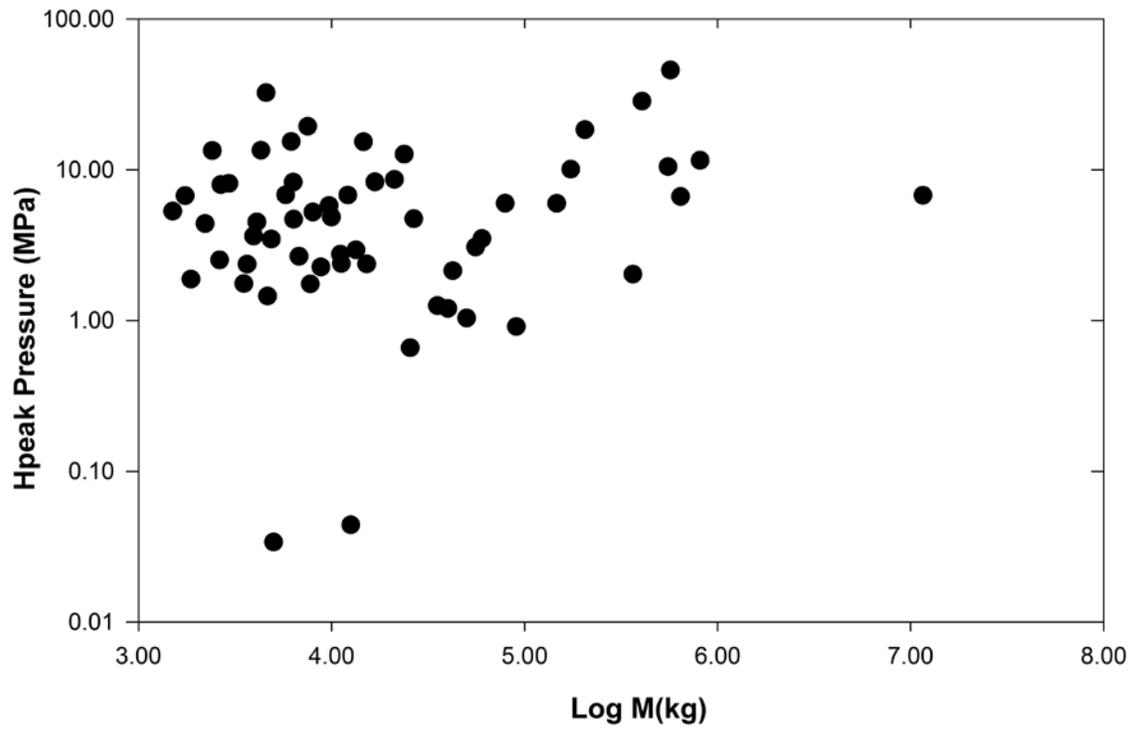

Figure 6.

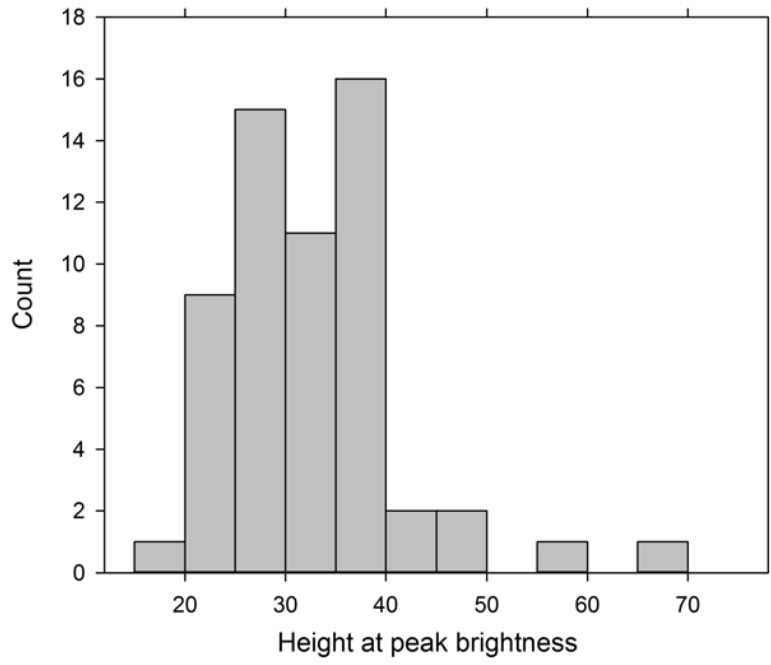

Figure 7.